\documentclass[journal=tches,final]{iacrtrans}

\usepackage{amsmath}
\usepackage{circuitikz}
\usepackage{tikz}
\usepackage{mathtools}
\usetikzlibrary{decorations.pathreplacing}
\usetikzlibrary{chains, shapes, decorations, arrows, calc, arrows.meta,fit,positioning,decorations.pathreplacing, decorations.pathmorphing, tikzmark, shapes.geometric, matrix, automata}
\usepackage{subcaption}
\definecolor{CWIred}{RGB}{205,0,58}
\colorlet{joint}{brown!40!orange!60}

\usepackage{algpseudocode}

\usepackage[linesnumbered,ruled,noend]{algorithm2e}

\SetNlSty{}{}{}
\let\oldnl\nl
\newcommand\nonl{%
  \renewcommand{\nl}{\let\nl\oldnl}}

\usepackage{multirow}
\usepackage{multicol}
\usepackage{enumerate}
\usepackage{booktabs}
\usepackage{makecell}
\usepackage{todonotes}
\usepackage{enumitem}

\newif\ifrevmarks \revmarksfalse
\newcommand{\rev}[1]{\ifrevmarks{\color{blue}#1}\else#1\fi}

\author{Mingfei Yu\inst{1} \and Chenglu Jin\inst{2} \and Giovanni De Micheli\inst{3}}
\institute{
  Stanford University, Stanford, CA, USA, \email{mfyu@stanford.edu}
  \and
  Centrum Wiskunde \& Informatica, Amsterdam, The Netherlands, \email{chenglu.jin@cwi.nl}
  \and
  EPFL, Lausanne, Switzerland, \email{giovanni.demicheli@epfl.ch}
}

\title[Efficient Secure Multi-Party Computation via Maximizing Local Processing]{Think Locally Before Computing Jointly:\\ Efficient Secure Multi-Party Computation via\\ Maximizing Local Processing}

\begin{document}

\maketitle


\begin{abstract}
    \emph{Secure multi-party computation}~(MPC) enables parties to jointly evaluate a function over private inputs without revealing them.
    A widely used MPC technique is the \emph{garbled circuit}~(GC) protocol, where performance is dominated by the amount of computation that must be executed cryptographically through the secure protocol.
    Prior GC optimization has largely equated this cost with the number of non-linear gates, especially AND gates, and has therefore focused on minimizing the multiplicative complexity of logic networks.
    In this work, we show that this abstraction misses an orthogonal source of efficiency: some gates, including non-linear ones, need not be garbled at all.
    If the inputs to a gate are known to, or locally computable by, one party, the gate can be evaluated outside the GC protocol without cryptographic overhead.

    Motivated by this observation, we introduce \emph{ownership-aware} optimization for GC-based MPC.
    Rather than minimizing only the total number of non-linear gates, our objective is to minimize the number of non-linear gates that require \emph{joint} secure evaluation.
    We formalize this objective through an ownership-aware cost model and build a compiler-integrated, gate-level optimization framework that automatically restructures Boolean circuits to maximize party-local computation while preserving secure computation semantics.
    Unlike application-specific rewrites, our approach is automated and general-purpose.
    Crucially, joint-computation reduction is orthogonal to the multiplicative-complexity minimization that underlies essentially all prior low-cost GC synthesis: it is a new dimension of the design space rather than a better AND-count heuristic, and it composes with existing techniques.
    Experimental results show that ownership-aware optimization exposes protocol-cost reductions missed by prior GC cost models, reducing jointly evaluated computation and improving end-to-end GC performance across circuits with party-local structure, opening a new perspective on low-cost garbled circuit design.
\end{abstract}

\section{Introduction}

\emph{Secure multi-party computation}~(MPC) enables mutually distrustful parties to jointly compute functions on private inputs without revealing them.
Among general-purpose MPC protocols, the \emph{garbled circuit}~(GC) paradigm, originating from Yao's seminal work on secure two-party computation~\cite{yao1986gc}, is particularly appealing due to its constant-round complexity and compatibility with arbitrary Boolean functions.
Over the past decades, a long line of protocol and implementation advances has substantially improved the efficiency of GC-based secure computation.
Representative milestones include garbled-row reduction~\cite{naor1999grr3}, the Free-XOR technique~\cite{kolesnikov2008freexor}, fixed-key garbling~\cite{bellare2013fixkey}, and the Three-Halves construction~\cite{rosulek2021threehalves}.
Together, these developments have shaped the modern cost profile of GC protocols: XOR gates can often be evaluated at essentially no cryptographic cost, whereas non-linear gates, especially AND gates, remain the primary source of cryptographic cost.

This cost structure has naturally motivated a synthesis-oriented line of work on \emph{optimal garbled circuit generation}.
Given a target Boolean function, these methods seek circuit implementations over gate libraries such as XOR, NOT, and AND, while minimizing the number of AND gates.
The underlying objective is closely related to \emph{multiplicative complexity}~\cite{boyar2000mc}: the minimum number of AND gates required to realize a Boolean function over the XOR-AND basis.
By reducing the number of non-linear gates, such techniques directly reduce the amount of computation that must be handled by the GC protocol~\cite{riazi2019mpcircuits}.

However, AND-count minimization relies on an implicit assumption: every non-linear gate appearing in the circuit contributes to cryptographic cost.
In this work, we revisit this assumption and observe that the cryptographic cost of a gate depends not only on its Boolean operator, but also on \emph{where its input values are available}.
A gate whose inputs are known to, or locally computable by, one party can be evaluated locally, even if the gate itself is non-linear.
Such a gate does not need to enter the GC protocol and therefore avoids the cryptographic cost of secure evaluation~\cite{volgushev2019conclave}.
In contrast, a gate whose evaluation depends on private information from multiple parties requires \emph{joint computation} and must be evaluated through the cryptographic mechanisms for a privacy-preserving evaluation.
This distinction motivates a refined optimization question: given a target function, how can one systematically construct a garbled circuit that minimizes the number of non-linear gates requiring joint computation, rather than merely minimizing the total number of non-linear gates?

Recent MPC systems reduce joint computation by decomposing the target computation and maximizing the portions executed locally.
In the domain of \emph{secure analytical SQL queries}, \texttt{Senate}~\cite{poddar2021senate} hand-designs Boolean-circuit implementations of SQL primitives (e.g., \texttt{filtering}, \texttt{joins}, \texttt{set intersection}) that minimize joint computation and manually composes them into query-specific circuits.
In secure machine-learning inference, \texttt{Marill}~\cite{rathee2024marill} applies \emph{layer freezing} during MPC-based fine-tuning of \emph{large language models}~(LLMs), avoiding joint computation for unchanged layers at a small accuracy cost.
Both achieve substantial gains, but their optimizations are manually engineered and tailored to a particular application domain, requiring significant expertise and not generalizing to arbitrary Boolean computations.
A separate line of automated compilers and languages for secure computation~\cite{kerschbaum2011automatically,rastogi2013knowledge,rastogi2014wysteria,acay2021viaduct,sweet2023symphony} instead reasons about what each party already knows to decide which computations can be performed locally; these systems, however, operate at the granularity of whole program operations and delegate the remaining secure computation to a separately optimized backend, leaving the joint computation \emph{within} the circuit untouched.
In this work, we pursue a complementary and more broadly applicable direction: an automated, gate-level optimization framework for garbled circuits that minimizes joint computation by systematically maximizing local computation.
Rather than relying on application-specific insights, our approach operates directly on Boolean circuit structure.
We develop combinatorial analysis and transformation techniques that identify sub-circuits amenable to single-party evaluation and restructure the circuit accordingly, thereby reducing the number of gates requiring joint evaluation without altering the target functionality.

We evaluate our framework on 12 Boolean circuits spanning arithmetic, floating-point, cryptographic, and general-purpose workloads, all preprocessed with high-effort MC-reduction techniques.
Our results show that ownership-aware optimization uncovers cryptographic-cost reductions that remain invisible to conventional AND-count minimization:
on a processor-style benchmark with substantial party-local control logic, our framework achieves a $39.25\%$ reduction in garbling cost;
in some cases, jointly evaluated computation decreases even when MC remains unchanged or slightly increases.
When integrated with \texttt{MP-SPDZ}~\cite{mp-spdz}, a widely used framework for implementing MPC protocols, these circuit-level reductions translate into concrete system-level savings.
These results confirm that ownership-aware optimization is complementary to existing garbled circuits synthesis techniques and can reduce the cryptographic cost of secure computation beyond what is captured by multiplicative complexity alone.
In doing so, this work introduces a new dimension into low-cost garbled circuit design that is orthogonal to the multiplicative-complexity minimization underlying essentially all prior synthesis efforts: rather than counting non-linear gates, we ask which of them must be evaluated jointly at all, opening a new perspective on secure computation efficiency that we believe can inform the next generation of design philosophy for MPC circuits.

The remainder of the paper is organized as follows.
\S\ref{sec:prelim} reviews the cryptographic and logic-synthesis foundations of our work.
\rev{\S\ref{sec:related} reviews related research topics}.
\S\ref{sec:overview} introduces our refined cost model based on \emph{ownership analysis} and presents the architecture of the circuit optimization framework.
\S\ref{sec:synthesis_engine} details the design of our partial circuit resynthesis engine.
\S\ref{sec:eval} reports an extensive experimental evaluation on diverse benchmarks, demonstrating the practical benefits of the proposed technique.
\rev{\S\ref{sec:position} positions our contributions relative to prior art.}
Finally, \S\ref{sec:conclusion} concludes the paper.
\section{Preliminaries}
\label{sec:prelim}

This section reviews the cryptographic and logic-synthesis background needed for our work.
We first summarize the GC protocol, its main cost components, and the threat model considered in this paper.
We then introduce the logic-network concepts that underlie our gate-level optimization framework, including Boolean circuit representations, cuts, multiplicative complexity, and Boolean decomposition.

\subsection{Cryptographic Preliminaries}
\label{subsec:crypto_prelim}

\subsubsection{Garbled Circuits}
\label{subsubsec:gc}

GC is a foundational cryptographic primitive enabling \emph{secure two-party computation}~(2PC) in the \emph{semi-honest} model.
In the protocol, two parties, commonly referred to as the \emph{garbler} and the \emph{evaluator}, jointly compute a Boolean function
\begin{math}
    f: \{0,1\}^n\mapsto \{0,1\}^m,
\end{math}
over their private inputs without revealing them to each other.
The function $f$ is typically represented as a Boolean circuit composed of two-input logic gates.

At a high level, the garbler assigns to each wire $w$ two random labels $W_w^0, W_w^1\in\{0,1\}^k$ encoding the bit values $0$ and $1$, where $k$ is the security parameter\rev{; under the \emph{free-XOR} optimization~\cite{kolesnikov2008freexor}, the two labels of a wire differ by a global secret offset $\Delta\in\{0,1\}^k$ known only to the garbler, i.e., $W_w^1=W_w^0\oplus\Delta$}.
For each two-input gate $g$ with inputs $u,v$ and output $w$, it builds a \emph{garbled truth table} of four ciphertexts: for each $(a,b)\in\{0,1\}^2$, the output label $W_w^{g(a,b)}$ is encrypted under the input labels $W_u^a$ and $W_v^b$, using symmetric-key primitives~\cite{yao1986gc,bellare2012gc}.
Applying this to all gates yields the \emph{garbled circuit}, which the garbler sends to the evaluator.

\begin{figure}
  \centering
  \resizebox{0.9\textwidth}{!}{\usetikzlibrary{shapes.gates.logic.US, calc}
\begin{tikzpicture}[scale=0.8, every node/.style={scale=0.8}]
    \tikzstyle{pi}  =[draw,minimum width=0.4cm, minimum height=0.4cm]
    \tikzstyle{gate}=[draw,minimum width=1cm]
    \node[gate, and gate US] at (0, 1) (and) {AND};
    \node at ($(and.input 1) + (-1, 0)$) (x) {};
    \node at ($(x) + (0, 0.25)$) (x_lab) {$x\in\{0, 1\}$};
    \node at ($(and.input 2) + (-1, 0)$) (y) {};
    \node at ($(y) + (0.5, -0.25)$) (y_lab) {$y$};
    \node at ($(and.output) + (1, 0)$) (z) {};
    \node at ($(z) + (-0.5, 0.25)$) (z_lab) {$z$};
    \draw (x.east) -- (and.input 1);
    \draw (y.east) -- (and.input 2);
    \draw (and.output) -- (z.west);
    \node at (-0.5, -0.5) (title1) {Boolean circuit};

    \node [shape=rectangle, align=center] (lut) at (8,1) {
        \begin{tabular}{|c|}
            \hline
            Enc$_{X^1, Y^1}$($Z^1$) \\
            \hline
            Enc$_{X^1, Y^0}$($Z^0$) \\
            \hline
            Enc$_{X^0, Y^1}$($Z^0$) \\
            \hline
            Enc$_{X^0, Y^0}$($Z^0$) \\
            \hline
        \end{tabular}
    };
    \node at ($(lut) + (-2.55, 0.25)$) (x_source) {};
    \node at ($(lut) + (-1.22, 0.25)$) (x_sink) {};
    \node at ($(x_source) + (-0.92, 0.25)$) (x_wire) {$X\in\{X^0, X^1\}\subseteq\{0, 1\}^k$};
    \node at ($(lut) + (-2.55, -0.25)$) (y_source) {};
    \node at ($(lut) + (-1.22, -0.25)$) (y_sink) {};
    \node at ($(y_source) + (0.7, -0.25)$) (y_wire) {$Y$};
    \node at ($(lut) + (1.05, 0)$) (z_source) {};
    \node at ($(lut) + (2.25, 0)$) (z_sink) {};
    \node at ($(z_sink) + (-0.5, 0.25)$) (z_wire) {$Z$};
    \node at (6.2, -0.5) (title2) {Garbled circuit};
    
    \draw (x_source) -- (x_sink);
    \draw (y_source) -- (y_sink);
    \draw (z_source) -- (z_sink);

    \node at (3.5, 0.3) (arr) {$\xRightarrow{\text{\normalsize garbling}}$};
\end{tikzpicture}}
  \caption{From an AND gate to its four-row garbled truth table.}
  \label{fig:garbling_and}
\end{figure}
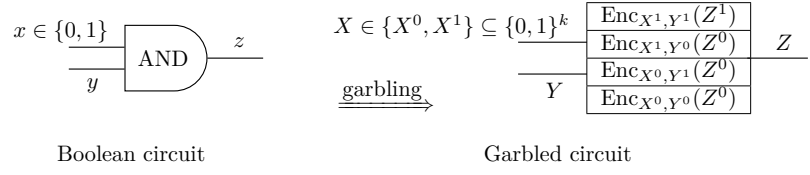

The evaluator obtains the labels for its own inputs via an \emph{oblivious transfer}~(OT) protocol and those of the garbler's inputs directly, then processes the circuit gate by gate in topological order: holding one label per input wire, it decrypts the matching row of each garbled table to recover the output label, which feeds subsequent gates.
Fig.~\ref{fig:garbling_and} illustrates this for a two-input AND gate ($z=x\wedge y$), whose four-row table encrypts each output label $Z^{x\wedge y}$ under the corresponding input labels $X^x$ and $Y^y$.
Throughout, the evaluator manipulates only opaque labels and learns nothing about intermediate values beyond the prescribed output, which is finally decoded using information provided by the garbler.

\subsubsection{Garbling Cost}
\label{subsubsec:garbling_cost}

We now summarize the main cost components of GC-based MPC.
Throughout this paper, we use the terms \emph{cryptographic cost} and \emph{garbling cost} interchangeably to refer to the communication and computation overhead incurred when operations are evaluated through the GC protocol.

\paragraph{Communication cost}
Communication arises primarily from two sources: the garbler transmits garbled tables to the evaluator, and the evaluator obtains its input wire labels through OT.

\paragraph{Computation cost}
Computation is dominated by the cryptographic operations to garble and evaluate non-linear gates, especially AND gates, each requiring symmetric-key operations and garbled-table transmission; under free-XOR~\cite{kolesnikov2008freexor}, XOR gates need no garbled tables and negligible computation.
Accordingly, the cryptographic cost of a garbled circuit is closely tied to the number of non-linear gates evaluated through the GC protocol.

\paragraph{Role of OT overhead}
OT cost scales with the number of evaluator input bits, so an implementation that exposes more evaluator inputs may increase OT communication and runtime.
In modern systems, however, OT extension~\cite{otextension} makes the marginal cost of additional inputs small: a fixed-cost setup phase (Base OT and correlation generation) dominates, while each extra input label adds little.
As our end-to-end evaluation shows (\S\ref{subsec:system_level_eval}), even large increases in evaluator input size yield only modest changes in OT cost, and recent input-label delivery mechanisms~\cite{dietz2025tinylabels,jin2024pg} reduce it further.
By contrast, each non-linear gate is a direct and persistent cost: with \emph{half-gates}~\cite{zahur2015halfgate}, every AND gate incurs fixed communication and computation while XOR remains free, so reducing non-linear gates deterministically cuts the work performed by the GC protocol.

These observations justify the common focus on minimizing non-linear gates in garbled-circuit generation, while also highlighting a subtlety that becomes central in our setting:
Our optimization may change the interface between \emph{local} computation and \emph{joint} computation, potentially altering the number of evaluator input labels.
Nevertheless, as we will show, reducing \emph{jointly evaluated non-linear gates} remains the dominant factor in reducing cryptographic cost.

\subsubsection{Threat Model}

This work assumes the \emph{semi-honest} adversarial model, wherein all parties follow the prescribed protocol honestly but may attempt to infer additional information from the data they receive during execution.
In the context of the GC protocol, this means that the garbler and evaluator execute their respective roles correctly, but may passively analyze wire labels, garbled tables, and protocol messages in an attempt to learn information about the other party's input.
Under standard assumptions, the GC protocol with common optimizations such as free-XOR, half-gates, and row reduction~\cite{naor1999grr3} is secure in this model: wire labels reveal no information about their underlying bit values, and the evaluator learns only the prescribed output.

Our work focuses on generating and optimizing garbled circuits, rather than designing a complete secure computation protocol.
Accordingly, the security guarantee we require is that the optimized garbled circuit preserves the same Boolean functionality and exposes no additional information beyond that revealed by the surrounding protocol.
The optimization does not alter the cryptographic mechanisms used to garble, transmit, or evaluate the circuit.

The optimization is therefore compatible with stronger (e.g., malicious) adversarial models when embedded in an appropriate secure protocol: maliciously secure systems such as \texttt{Senate}~\cite{poddar2021senate} could use optimized garbled circuits provided the protocol enforces the necessary checks, such as \emph{input consistency} and \emph{correctness of garbling}; designing such protocols is beyond our scope.

\subsection{Logic Synthesis and Optimization}

\subsubsection{Logic Networks}


This work adopts logic networks as the primary representation of Boolean functions.
Formally, a logic network is a \emph{directed acyclic graph}~(DAG) $G=(V,E)$, where each node $v\in V$ represents either a \emph{primary input}~(PI) or a Boolean operator (e.g., AND, XOR), and each edge $(u,v)\in E$ denotes a logical dependency -- the output of node $u$ serves as an input to node $v$.
A PI node has no fan-in; all other nodes have fixed fan-in and compute a Boolean function of their immediate predecessors.
\emph{Primary outputs}~(POs) are not treated as distinct node types, but rather as tags applied to existing nodes.
Each PO tag specifies a node whose value is designated as an output of the network, optionally with a logic complementation applied as an attribute of the tag.

A key structural concept is the \emph{cut}: for a node $v$ (the \emph{root}), a cut is a set of nodes $L\subseteq V$ -- the \emph{leaves} -- such that every path from a primary input to $v$ passes through some node in $L$.
Written $c=(L,v)$, it represents the subcircuit rooted at $v$ with $L$ as inputs, abstracting $v$'s logic cone into a self-contained sub-function that can be isolated for resynthesis -- a foundational primitive in modern synthesis flows.
We focus on $\kappa$-feasible cuts ($|L|\leq\kappa$), balancing optimization granularity and computational efficiency.

We restrict logic networks to the gate library \{AND, XOR, NOT\}: under the free-XOR technique, XOR and NOT are free while AND incurs cryptographic cost, so this basis captures the linear/non-linear asymmetry central to garbling.
Modeling NOT as an edge attribute yields the compact \emph{XOR-AND-inverter graph}~(XAG), widely supported in synthesis and GC frameworks.
\rev{Concretely, an XAG contains only AND and XOR nodes; a NOT is never materialized as a node but recorded as a complemented edge.
This mirrors garbling practice: a wire $w$ carries the label pair $\{W_w, W_w\oplus\Delta\}$ regardless of polarity, so the garbler realizes a NOT by swapping which label encodes $0$ and $1$ when garbling the consuming gate, while the evaluator does nothing; consequently, NOT gates neither contribute to MC nor to JMC, and ownership propagates through a complemented edge unchanged (defined later in \S\ref{subsec:ownership}).}
Although specialized representations for garbled-circuit optimization exist~\cite{yu2023x1gopt}, we adopt XAGs for toolchain compatibility, and our framework could be extended to such structures as needed.

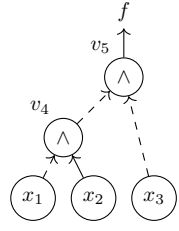
\begin{figure}[t]
    \centering
    \begin{tikzpicture}[scale=0.8, every node/.style={scale=0.8}]
    \node[state, xshift=0cm, yshift=0cm, minimum size=0.5cm] (a1) {$x_1$};
    \node[state, xshift=1cm, yshift=0cm, minimum size=0.5cm] (b1) {$x_2$};
    \node[state, xshift=2cm, yshift=0cm, minimum size=0.5cm] (b2) {$x_3$};

    \node[anchor=base] (fo) at (1.5,3) {$f$};
    
    \node[state, xshift=0.5cm, yshift=1cm, minimum size=0.5cm] (v4) {$\land$};
    \node[anchor=base] (v4_i) at (0.115,1.45) {$v_4$};
    
    \node[state, xshift=1.5cm, yshift=2cm, minimum size=0.5cm] (v5) {$\land$};
    \node[anchor=base] (v5_i) at (1.115,2.45) {$v_5$};
    
    \draw [->, dashed] (a1) edge (v4);
    \draw [->] (b1) edge (v4);
    \draw [->, dashed] (v4) edge (v5);
    \draw [->, dashed] (b2) edge (v5);
    \draw [->] (v5) edge (fo);
\end{tikzpicture}
    \caption{A logic network implementing function \texttt{\#0b}.}
    \label{fig:logic_network_example}
\end{figure}

An example logic network for a three-variable function $f$ (truth table \texttt{\#0b}~\footnote{We represent a truth table as a bit-string in hexadecimal, with the most significant bit on the left-hand side.}) is shown in Fig.~\ref{fig:logic_network_example}: three PI nodes $x_1,x_2,x_3$ and two internal AND nodes $v_4,v_5$, with dashed edges denoting negation and $v_5$ tagged as the PO.
It also illustrates cuts: the root $v_5$ has two $3$-feasible cuts, with leaf sets $\{x_3,v_4\}$ and $\{x_1,x_2,x_3\}$, each sufficient to compute $v_5$ and thus forming a functional boundary for local optimization.
Although this network contains no XOR gates, it still qualifies as an XAG by definition.

\subsubsection{Multiplicative Complexity}

The \emph{multiplicative complexity}~(MC) of a Boolean function was originally introduced as a measure of its inherent non-linearity.  
Formally, the MC of a Boolean function is defined as the minimum number of AND gates required to compute it over the basis $\{\mathrm{AND}, \mathrm{XOR}, \mathrm{NOT}\}$~\cite{boyar2000mc}.  
This is a \emph{functional} property -- it depends solely on the function’s truth table, independent of any specific implementation.

In the context of garbled circuit generation, we are often concerned with a \emph{structural} variant of this measure: the number of AND gates present in a particular implementation of the function as an XAG.  
This \emph{structural MC} directly corresponds to the number of non-linear gates that must be garbled~\cite{yu2023x1g}, and thus serves as a natural optimization objective for reducing the communication and computation costs of GC protocols.

Let $G = (V, E)$ be an XAG with PI set $I \subseteq V$, each internal node $v \in V \setminus I$ being labeled with an operator $\circ_v \in \{\wedge, \oplus\}$; the structural MC of $G$ is $\mathsf{MC}(G) = |\{ v \in V \setminus I \mid \circ_v = \wedge \}|$\footnote{Unless otherwise specified, MC in this report refers to structural MC.} (two for Fig.~\ref{fig:logic_network_example}).
This metric has been the dominant cost proxy in prior GC optimization research~\cite{riazi2019mpcircuits}.

Functional MC lower-bounds the structural MC of any XAG implementation, but determining it is computationally hard: except for special classes (e.g., symmetric functions~\cite{boyar2008mcsym}), exact methods are limited to functions with at most six inputs~\cite{calik2019mc6}.
Practical MC minimization therefore relies on heuristic synthesis, seeking an XAG with as few AND gates as possible~\cite{testa2020mc}.

\subsubsection{Boolean Decomposition}
\label{subsubsec:Boolean_decomposition}

Boolean decomposition breaks a function into simpler sub-functions that recombine to the original.
This divide-and-conquer strategy enables scalable synthesis by applying advanced techniques -- e.g., SAT-based exact synthesis~\cite{haaswijk2020sat} -- to smaller, tractable components.



There exists many Boolean decompsition techniques in the literature, each suited to certain optimization objectives.
In this work, we focus on \emph{Ashenhurst–Curtis decomposition}~(ACD)~\cite{ashenhurst1959,curtis1962switch}, which provides a general and expressive framework for partitioning input variables and restructuring Boolean functions.

ACD partitions the input variables of a Boolean function into three pairwise disjoint subsets: the \emph{bound set}~(BS), \emph{shared set}~(SS), and \emph{free set}~(FS), enabling structured factorization of the function.
For a single-output Boolean function $f(\vec{x})$, the ACD expresses the function in the form:
\begin{equation*}
    f(\vec{x})=l\big(\vec{g}(\vec{x}_{\text{BS}},\ \vec{x}_{\text{SS}}),\ \vec{x}_{\text{SS}},\,\vec{h}(\vec{x}_{\text{FS}})\big),
\end{equation*}
where $\vec{x} = \vec{x}_{\text{BS}} \cup \vec{x}_{\text{SS}} \cup \vec{x}_{\text{FS}}$ and the subsets are pairwise disjoint.
The sub-functions $\vec{g}$ and $\vec{h}$ are commonly referred to as the BS and FS functions, respectively, while $l$ denotes the combining logic.

Unlike variable-based decompositions, ACD operates on partitions of the input space, enabling more expressive structural transformations.
It has been widely used in the synthesis of lookup-table (LUT) networks~\cite{mishchenko2008decomp,calvino2025decomp}, a standard representation in \emph{field programmable gate array}~(FPGA) design automation.
\section{Related Works}
\label{sec:related}

\subsection{Cryptographic Optimizations for GC}
\label{subsec:crypto}

A series of cryptographic optimizations shapes the modern GC cost model, establishing a strong asymmetry in which linear gates are essentially free while non-linear gates dominate communication and computation.
The \emph{free-XOR} technique~\cite{kolesnikov2008freexor} evaluates XOR gates without any cryptographic operation or communication, while a line of constructions steadily reduces the cost of each AND gate: \emph{garbled row reduction}~\cite{naor1999grr3}, \emph{half-Gates}~\cite{zahur2015halfgate} as combined in modern implementations such as \texttt{MP-SPDZ}~\cite{mp-spdz}, and the more recent \emph{three-halves} construction~\cite{rosulek2021threehalves}.
Our work introduces no new cryptographic construction; it improves efficiency at the circuit level under the cost model these optimizations establish, and is therefore complementary to -- and naturally combinable with -- this line of research.

\subsection{Optimal Garbled Circuit Generation}

Prior works on automatically generating low-cost garbled circuits have primarily framed the problem as one of \emph{multiplicative complexity}~(MC) reduction, borrowing \emph{electronic design automation}~(EDA) techniques: with asymmetric gate costs reflecting free-XOR, the task reduces to finding XAG implementations of the target function with as few AND gates as possible.

This paradigm was notably advanced by the \texttt{TinyGarble} framework~\cite{songhori2015tinygarble}, which casts garbled circuit generation as a logic synthesis problem and leverages standard hardware synthesis tools with customized cost models to minimize non-XOR gates.
Beyond combinational optimization, \texttt{TinyGarble} further introduced the use of sequential circuit representations to improve scalability and reduce memory footprint, enabling efficient handling of large functions.

In parallel, a series of works have explored logic optimization techniques tailored for MC reduction, including cut-based \emph{rewriting}, \emph{resubstitution}, and exploitation of \emph{structural don't-cares}~\cite{testa2019mc,testa2020mc,liu2022dontcare}.
Collectively, these approaches aim to minimize the number of AND gates under the assumption that each non-linear gate incurs identical cost in secure evaluation.

Beyond dedicated MC-reduction heuristics, a complementary line of work repurposes general-purpose logic synthesis and hardware design automation for secure computation.
Demmler et al.~\cite{demmler2015automated} adapt industrial logic-synthesis tools with custom technology-mapping libraries whose cost metric targets MPC -- for example, minimizing non-linear gates or circuit depth -- rather than silicon area or delay.
B\"uscher et al.~\cite{buscher2017compiling} compile ANSI~C programs into Boolean circuits minimized for MPC cost, combining model-checking-based circuit generation with gate-level minimization.
\texttt{ARM2GC}~\cite{songhori2019arm2gc} instead garbles an entire ARM processor so any program compiled to that instruction set can be securely evaluated, and its \texttt{SkipGate} algorithm skips garbling and communication for gates publicly determined by the instruction stream.
Like the MC-reduction paradigm, these works assume every non-linear gate requiring secure evaluation is equally costly; our ownership-aware model refines this by distinguishing gates that can be evaluated locally.


\subsection{Exploiting Party Knowledge in Secure Computation}
\label{subsec:party_knowledge}

A separate line of work reduces secure-computation cost by reasoning about what each party already knows, moving computation out of the cryptographic protocol whenever a value can be derived locally.
Kerschbaum~\cite{kerschbaum2011automatically} combines static program analysis with a theorem prover to identify intermediate values a party can already determine -- from its own inputs together with the revealed output -- and evaluates them outside the secure protocol;
Urmoneit~\cite{urmoneit2013efficient} replaces the theorem prover with a dedicated optimization algorithm to scale to larger programs.
Rastogi et al.~\cite{rastogi2013knowledge} generalize this through \emph{knowledge inference}, revealing and locally computing an intermediate value when a party gains no more from observing it than from the intended output.
Notably, this line of work exploits information disclosed by the \emph{output}, and not only the parties' inputs.

The same principle underlies several secure-computation programming languages and compilers.
\texttt{Wysteria}~\cite{rastogi2014wysteria} provides a verified language for \emph{mixed-mode} computation, letting programmers interleave single-party local computation with secure multiparty blocks.
\texttt{Viaduct}~\cite{acay2021viaduct} compiles a single information-flow-labeled program into a distributed protocol, placing each computation either in the clear on an authorized host or under a cryptographic backend as the security labels permit.
\texttt{Symphony}~\cite{sweet2023symphony} offers first-class shares and party sets, enabling cleartext and MPC operations to be mixed and coordinated reactively across parties.

These approaches decide at the granularity of whole program operations which execute locally versus securely, and treat the secure portion as a black box optimized separately, e.g., by MC minimization.
Our work is complementary but gate-level: we restructure the Boolean circuit so that individual non-linear gates become locally evaluable, reducing the joint computation \emph{within} the secure portion these compilers would otherwise emit.
Moreover, whereas some of these works additionally exploit knowledge implied by the output~\cite{kerschbaum2011automatically,rastogi2013knowledge}, our current framework leverages only input ownership; extending ownership analysis to incorporate output-derived knowledge is a promising direction for further reducing joint computation.

\subsection{Joint Computation Minimization}

This work is not the first attempt to reduce the amount of joint computation in MPC.
As discussed in the introduction, prior efforts have explored this idea in application-specific contexts.  
For example, \texttt{Senate}~\cite{poddar2021senate} demonstrates that carefully designed circuits can minimize joint computation for analytical SQL queries.
Similarly, recent works in privacy-preserving machine learning show that reducing the extent of joint computation during fine-tuning -- for instance, by leveraging layer freezing in secure \emph{large language model}~(LLM) inference~\cite{rathee2024marill} -- can significantly improve efficiency.

Our contribution, however, is fundamentally different in scope.
To the best of our knowledge, this is the first \emph{automated}, \emph{gate-level} framework for \emph{general-purpose} Boolean garbled circuit-based MPC that directly targets joint computation reduction: in contrast to the program-level compilers of \S\ref{subsec:party_knowledge}, which decide locality at the granularity of whole operations, we reduce joint computation by restructuring the circuit at the level
of individual gates, turning manual, application-specific optimizations into a generic and scalable methodology.
\section{Overview}
\label{sec:overview}

This section presents a high-level overview of our proposed optimization framework for garbled circuit optimization via joint computation minimization.
We begin by introducing a refined cost model based on the notion of \emph{node ownership}, which distinguishes between operations that require secure joint evaluation and those that can be executed locally. We then outline the architecture of our optimization framework, highlighting its cut-based peephole optimization strategy.
Finally, we discuss the role of Boolean decomposition in enabling efficient resynthesis, and motivate the use of ACD as the backbone of our resynthesis engine.

\subsection{Node Ownership and Refined Cost Model}
\label{subsec:ownership}

\subsubsection{Limitations of MC as a Cost Proxy}

While MC is a coarse indicator of garbling cost, it does not capture the actual cost incurred in GC-based MPC.
As observed in prior work (e.g., \cite{volgushev2019conclave}), garbled gates whose execution depends solely on one party's data need not be part of the MPC protocol and can be evaluated locally in plaintext, free of cryptographic overhead.
Cost is therefore dictated not by the total number of AND nodes in the XAG but by the subset whose evaluation requires joint inputs; using MC as the sole proxy overestimates cost by failing to distinguish local from joint computation.

Beyond passively excluding locally derivable operations -- itself a near cost-free gain -- we ask whether one can \emph{actively} restructure the logic to reduce the number of jointly evaluated operations, possibly at the expense of increased local processing.
To formalize this, we introduce the notion of node ownership, which statically determines whether a node in a logic network requires joint computation or can be evaluated locally, yielding a cost model that separates the truly costly components of an MPC task from those offloadable without interaction.

\subsubsection{Node Ownership}
\label{subsubsec:node_ownership}

To enable finer-grained cost modeling for GC–based MPC, we introduce the notion of node ownership, which classifies each node in a logic network as either \emph{locally derivable} or \emph{jointly derived}.
This distinction reflects whether the node's computation depends solely on inputs from a single party or requires data from multiple parties and must therefore be computed securely within the MPC protocol.

Let $G=(V,E)$ be a logic network with PI nodes $I\subseteq V$, and let $\mathcal{P}=\{\mathcal{P}_1,\mathcal{P}_2,\dots,\mathcal{P}_t\}$ denote the set of $t$ semi-honest parties participating in the computation.
Each PI node $v\in I$ is owned by exactly one party, indicated by a mapping, $\mathsf{own}: I\mapsto \mathcal{P},$ which specifies the party $\mathcal{P}_i\in\mathcal{P}$ that privately holds the value of $v$.

We define the set of locally derivable nodes with respect to party $\mathcal{P}_i$, denoted $\mathsf{LD}_{\mathcal{P}_i}\subseteq V$, recursively as:
\begin{equation*}
    v \in \mathsf{LD}_{\mathcal{P}_i}
    \iff
    \begin{cases}
        v \in I \ \text{and} \ \mathsf{own}(v) = \mathcal{P}_i, \quad\text{or}\\
        \forall u \in \mathrm{fanin}(v), \ u \in \mathsf{LD}_{\mathcal{P}_i}.
    \end{cases}
\end{equation*}
In other words, a node is locally derivable by party $\mathcal{P}_i$ if it is either: (1) a PI owned by $\mathcal{P}_i$, or (2) a gate whose fanin nodes are all locally derivable by $\mathcal{P}_i$.

The set of jointly derived nodes, denoted $\mathsf{JD} \subseteq V$, is defined as:
\begin{equation*}
    \mathsf{JD} = \{ v \in V \mid \forall i \in \{1,\dots,t\}, \ v \notin \mathsf{LD}_{\mathcal{P}_i} \}. 
\end{equation*}
That is, a node belongs to $\mathsf{JD}$ if it is not locally derivable with respect to any single party.

We further define the \emph{frontier set}, denoted $\mathsf{FR} \subseteq \mathsf{JD}$, as:
\begin{equation*}
    \mathsf{FR} = \{ v \in \mathsf{JD} \mid \exists u \in \mathrm{fanin}(v) \ \text{s.t.} \ u \in \mathsf{LD}_{\mathcal{P}_i} \ \text{for some} \ i \}.
\end{equation*}
Intuitively, the frontier consists of jointly derived nodes that are immediately preceded by at least one locally derivable node.  
This concept is of particular interest, since the goal of this work, namely, minimizing joint computation via maximizing local data processing, can be viewed as a process of ``pushing'' the frontier from the PI side towards the PO side.

\subsubsection{Refined Cost Model}

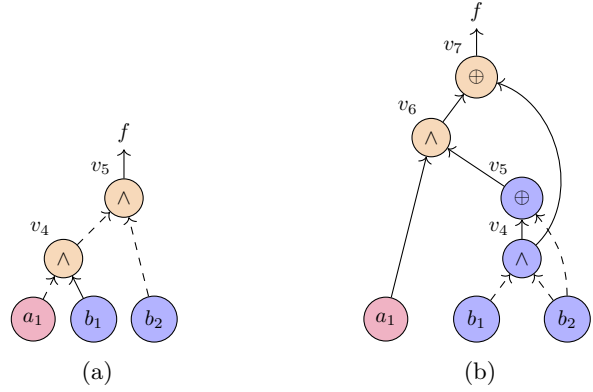
\begin{figure}[t]
    \centering
    \begin{subfigure}{.2\textwidth}
        \centering
        \begin{tikzpicture}[scale=0.8, every node/.style={scale=0.8}]
    \node[state, xshift=0cm, yshift=0cm, minimum size=0.5cm, fill=CWIred!30] (a1) {$a_1$};
    \node[state, xshift=1cm, yshift=0cm, minimum size=0.5cm, fill=blue!30] (b1) {$b_1$};
    \node[state, xshift=2cm, yshift=0cm, minimum size=0.5cm, fill=blue!30] (b2) {$b_2$};

    \node[anchor=base] (fo) at (1.5,3) {$f$};
    
    \node[state, xshift=0.5cm, yshift=1cm, minimum size=0.5cm, fill=joint!50] (v4) {$\land$};
    \node[anchor=base] (v4_i) at (0.115,1.45) {$v_4$};
    
    \node[state, xshift=1.5cm, yshift=2cm, minimum size=0.5cm, fill=joint!50] (v5) {$\land$};
    \node[anchor=base] (v5_i) at (1.115,2.45) {$v_5$};
    
    \draw [->, dashed] (a1) edge (v4);
    \draw [->] (b1) edge (v4);
    \draw [->, dashed] (v4) edge (v5);
    \draw [->, dashed] (b2) edge (v5);
    \draw [->] (v5) edge (fo);
\end{tikzpicture}
        \caption{}
        \label{fig:example_orig}
    \end{subfigure}
    \hspace{5em}
    \begin{subfigure}{.25\textwidth}
        \centering
        \begin{tikzpicture}[scale=0.8, every node/.style={scale=0.8}]
    \node[state, xshift=0cm, yshift=0cm, minimum size=0.5cm, fill=CWIred!30] (a1) {$a_1$};
    \node[state, xshift=1.5cm, yshift=0cm, minimum size=0.5cm, fill=blue!30] (b1) {$b_1$};
    \node[state, xshift=3cm, yshift=0cm, minimum size=0.5cm, fill=blue!30] (b2) {$b_2$};

    \node[anchor=base] (fo) at (1.5,5) {$f$};

    \node[state, xshift=2.25cm, yshift=1cm, minimum size=0.5cm, fill=blue!30] (v4) {$\land$};
    \node[anchor=base] (v4_i) at (1.865,1.45) {$v_4$};

    \node[state, xshift=2.25cm, yshift=2cm, minimum size=0.5cm, fill=blue!30] (v5) {$\oplus$};
    \node[anchor=base] (v5_i) at (1.865,2.45) {$v_5$};

    \node[state, xshift=0.75cm, yshift=3cm, minimum size=0.5cm, fill=joint!50] (v6) {$\land$};
    \node[anchor=base] (v6_i) at (0.365,3.45) {$v_6$};
    
    \node[state, xshift=1.5cm, yshift=4cm, minimum size=0.5cm, fill=joint!50] (v7) {$\oplus$};
    \node[anchor=base] (v7_i) at (1.115,4.45) {$v_7$};
    
    \draw [->, dashed] (b1) edge (v4);
    \draw [->, dashed] (b2) edge (v4);
    \draw [->] (v4) edge (v5);
    \draw [->, dashed] [bend right=20] (b2) edge (v5);
    \draw [->] (v5) edge (v6);
    \draw [->] (a1) edge (v6);
    \draw [->] (v6) edge (v7);
    \draw [->] [bend right=60] (v4) edge (v7);
    \draw [->] (v7) edge (fo);
\end{tikzpicture}
        \caption{}
        \label{fig:example_opt}
    \end{subfigure}
    \caption{Two MC-optimal XAG implementations of the Boolean function from Fig.~\ref{fig:logic_network_example} under the ownership assumption that $a_1$ is owned by party $\mathcal{A}$ and $b_1$, $b_2$ by party $\mathcal{B}$. Nodes locally derivable to $\mathcal{A}$ are colored \textcolor{CWIred!70}{red}, those locally derivable to $\mathcal{B}$ are colored \textcolor{blue!70}{blue}, and jointly derived nodes are colored \textcolor{joint}{orange}. (a) Both AND gates ($v_4$, $v_5$) are jointly derived, yielding $\mathrm{JMC}=2$. (b) Only one AND gate ($v_6$) is jointly derived, yielding $\mathrm{JMC}=1$, as $v_4$ can be processed locally by $\mathcal{B}$.}
    \label{fig:limitation_of_mc}
\end{figure}

The limitations of using MC alone as the cost proxy for garbled circuits generation can be illustrated by the example in Fig.~\ref{fig:limitation_of_mc}.
We revisit the Boolean function from Fig.~\ref{fig:logic_network_example} (truth table: \texttt{\#0b}) under a new MPC context: among the three PIs, $x_1$, $x_2$, and $x_3$, the first originates from party $\mathcal{A}$, while the latter two originate from party $\mathcal{B}$.
For clarity, we rename them as $a_1$, $b_1$, and $b_2$, respectively.

Both XAG implementations in Fig.~\ref{fig:limitation_of_mc} are MC-optimal: for this three-variable function, SAT-based exact synthesis~\cite{soeken2020determining} gives \emph{functional MC} $2$, so at least two AND gates are required, and both meet this bound.
Under MC as the sole metric they are equally good -- indeed Fig.~\ref{fig:limitation_of_mc}\subref{fig:example_orig} might even look preferable, having fewer total nodes.

However, incorporating the notion of node ownership reveals a different picture.
We annotate nodes locally derivable to $\mathcal{A}$ and $\mathcal{B}$ in red and blue, respectively, and jointly derived nodes in orange.
Although both implementations contain exactly two AND gates, in Fig.~\ref{fig:limitation_of_mc}\subref{fig:example_orig} both AND gates ($v_4$ and $v_5$) are jointly derived, each incurring one unit of garbling cost.
In contrast, in Fig.~\ref{fig:limitation_of_mc}\subref{fig:example_opt}, only one of the two AND gates ($v_6$) is jointly derived; the other ($v_4$) is locally derivable to $\mathcal{B}$ and can be evaluated offline without any MPC overhead.
Consequently, when adopting the implementation in Fig.~\ref{fig:limitation_of_mc}\subref{fig:example_opt} for garbled circuit generation, the effective inputs from $\mathcal{B}$ become $v_4$ and $v_5$ rather than $b_1$ and $b_2$.

This example shows that MC alone cannot identify computation executable outside the GC protocol; our refined cost model addresses this by focusing on the jointly derived AND nodes.
Formally, for an XAG $G=(V,E)$, let $\mathsf{JD} \subseteq V$ be the set of jointly derived nodes as defined in \S~\ref{subsubsec:node_ownership}.
We define the \emph{joint multiplicative complexity}~(JMC) as:
\begin{equation*}
    \mathsf{JMC}(G) = \left|\left\{ v \in \mathsf{JD}\ \middle|\ \circ_v = \wedge \right\}\right|.
\end{equation*}
The JMC precisely counts the number of AND gates whose evaluation requires joint computation, and thus serves as the target metric for our logic optimization framework.
By minimizing JMC rather than MC, we directly reduce the actual garbling cost and improve MPC efficiency.

\subsection{Logic Optimization Framework}

\begin{algorithm}[t]
    \DontPrintSemicolon
    \caption{Peephole Optimization Framework}
    \label{alg:frame}
    \KwIn{XAG $G$, synthesis engine $\mathsf{Syn}$, cut size bound $\kappa$}
    \KwOut{Optimized XAG $G$}
    \ForEach{$v \in V(G)$ in topological order}
    {
        $\mathcal{C} \gets \mathsf{EnumerateCuts}(G, v, \kappa)$\;
        $G_v^{\text{best}} \gets \texttt{NULL}$\;
        $jmc_v^{\text{best}} \gets \mathsf{JMC}(G)$\;
        \ForEach{$c \in \mathcal{C}$}
        {
            $f_c \gets \mathsf{ExtractFunction}(G, c)$\;
            $G_{c}^{\text{cand}} \gets \mathsf{Syn}(f_c)$\;
            \If{$\neg \mathsf{StructurallyDifferent}(G_{c}^{\text{cand}}, c)$}
            {
                \textbf{continue}
            }
            \If{$\mathsf{JMC}(G_{c}^{\text{cand}}) < \mathsf{JMC}(c)$}
            {
                $G_{\text{temp}} \gets \mathsf{Replace}(G, c, G_{c}^{\text{cand}})$\;
                \If{$\mathsf{JMC}(G_{\text{temp}}) < jmc_v^{\text{best}}$}
                {
                    $G_v^{\text{best}} \gets G_{\text{temp}}$\;
                    $jmc_v^{\text{best}} \gets \mathsf{JMC}(G_{\text{temp}})$\;
                }
            }
        }
        \If{$G_v^{\text{best}} \neq \texttt{NULL}$}
        {
            $G \gets G_v^{\text{best}}$\;
        }
    }
    \Return{$G$}\;
\end{algorithm}

Our optimization framework adopts a peephole rewriting strategy operating directly at the gate-netlist level of an XAG.
The core idea is to iteratively identify small sub-networks, represented by cuts of bounded size, and resynthesize them using a dedicated synthesis engine (introduced in \S\ref{sec:synthesis_engine}).
The objective is to minimize the number of jointly derived AND nodes (i.e., JMC) while preserving functional equivalence.
This strategy is inspired by \cite{riener2019dag}, but adapted to our cost model.

As presented in Algorithm~\ref{alg:frame}, the process begins with a topological traversal of all nodes in $G$.
For each node $v$, we enumerate all cuts rooted at $v$ whose number of leaves does not exceed the bound $\kappa$ (line 2).
We initialize placeholders $G_v^{\text{best}}$ and $jmc_v^{\text{best}}$ (lines 3–4) to store the best candidate implementation for node $v$.

Each cut $c$ defines a $\kappa_0$-input, single-output Boolean function $f_c$, where $\kappa_0\leq\kappa$.
Once $f_c$ is derived (line 6), the synthesis engine is invoked on the function to obtain a candidate implementation $G_{c}^{\text{cand}}$ (line 7).
Candidates structurally identical to the current implementation are skipped.

Each candidate is screened in two stages.
Stage 1 (line 10) compares the candidate's JMC against the current cut in isolation; if it improves, Stage 2 (line 12) performs a global dry run -- temporarily replacing $c$ with $G_{c}^{\text{cand}}$ and re-evaluating JMC over the whole network to account for shared sub-structures -- and records the candidate if it beats the best so far for $v$ (lines 13-14).
Only after all cuts of $v$ are examined is the best implementation committed (lines 15-16), so each root adopts the replacement yielding the largest global reduction, coupling aggressive local search with global caution.

\subsection{Role of Boolean Decomposition}

A key design choice is to base the synthesis engine on Boolean decomposition, motivated primarily by \emph{scalability}: decomposition-based methods handle larger functions, letting us explore larger cuts (higher $\kappa$) that capture the broader structural context needed to reduce joint computation.
Empirically, for joint computation minimization \emph{scope} matters more than \emph{local optimality}: larger cuts expose richer structural relationships and enable transformations inaccessible to small, locally optimized sub-circuits, so reasonably good implementations over larger regions consistently \rev{tend to} yield greater reductions in jointly evaluated computation \rev{(\S~\ref{par:kappa_sensitivity} quantifies this and its limits)}.

The effectiveness of decomposition, however, depends on how it is guided.
A structure-agnostic scheme such as Shannon decomposition, which expands functions in a uniform, variable-driven manner, ignores the target function's structure and may fail to align with minimizing joint computation.
\rev{Concretely, Shannon expansion~\cite{shannon1938switch} over the $|\vec{a}|$ variables owned by party $\mathcal{A}$ yields $2^{|\vec{a}|}$ cofactors free of $\vec{a}$, which $\mathcal{A}$ can pre-evaluate locally, plus a selection layer mixing both parties' inputs whose size grows exponentially in $|\vec{a}|$ regardless of the structure of $f$; Davio expansions~\cite{davio1978} are XOR-based variants with the same behavior.
Bi-decomposition~\cite{mishchenko2001bidecomp}, $f=h(g_1(\vec{x}_1),g_2(\vec{x}_2))$ with a fixed two-input $h$, splits the support into two blocks, but neither block corresponds to a party and the ownership of $g_1,g_2$ is not controllable.
\emph{Ashenhurst-Curtis decomposition}~(ACD), in contrast, is a functional decomposition $f=l(\vec{g}(\vec{x}_{\mathrm{BS}}),\vec{h}(\vec{x}_{\mathrm{FS}}))$ whose variable sets we choose: placing single-owner variables in the free set makes every $\vec{h}$ local by construction, while the number of jointly evaluated select signals, $|\vec{g}|=\lceil\log_2\mu\rceil$ for column multiplicity $\mu$, adapts to the structure of $f$ instead of growing with $|\vec{a}|$.
ACD is thus the classical decomposition whose degrees of freedom coincide with the ownership partition, which is why we selected it.}
This motivates a structure-aware strategy; in the next section we introduce our synthesis engine based on ACD, which enables targeted variable partitioning aligned with maximizing local computation while minimizing joint evaluation.

\section{Methodology}
\label{sec:synthesis_engine}

Building on the insights from the previous section, we now present the design of our Boolean synthesis engine tailored to minimize jointly derived computation.
At the core of our approach is an ACD-based strategy, which enables systematic partitioning of functions into sub-components aligned with party-specific input ownership.

\subsection{Joint Compute-Oriented Decomposition}

Recall the symbolic formulation of ACD introduced in \S~\ref{subsubsec:Boolean_decomposition} $f(\vec{x}) = l\big(\vec{g}(\vec{x}_{\text{BS}},\vec{x}_{\text{SS}}),\ \vec{x}_{\text{SS}}, \vec{h}(\vec{x}_{\text{FS}})\big),$ where $\vec{g}$ are the BS functions, $\vec{h}$ are the FS functions, and $l$ is a top function that, together, realize $f$.

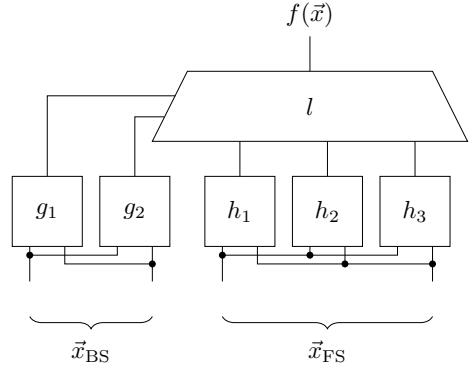
\begin{figure}[t]
    \centering
    \resizebox{0.45\linewidth}{!}{\begin{tikzpicture}
        \draw (2,4) -- (6.5,4) -- (6,5) -- (2.5,5) -- cycle;
        \node at (4.25,4.5) {$l$};
        \draw (4.25,5) -- (4.25,5.5) node[above] {$f(\vec{x})$};

        \draw (0,2.5) rectangle (1,3.5);
        \node at (0.5,3) {$g_1$};
        \draw (0.5,3.5) -- (0.5,4.65) -- (2.33,4.65);
        
        \draw (1.25,2.5) rectangle (2.25,3.5);
        \node at (1.75,3) {$g_2$};
        \draw (1.75,3.5) -- (1.75,4.35) -- (2.17,4.35);

        \draw (0.25,2) -- (0.25,2.5);
        \draw (2,2) -- (2,2.5);

        \draw (0.25,2.375) -- (1.5,2.375) -- (1.5,2.5);
        \node[fill=black, circle, inner sep=1pt] at (0.25,2.375) {};
        \draw (2,2.25) -- (0.75,2.25) -- (0.75,2.5);
        \node[fill=black, circle, inner sep=1pt] at (2,2.25) {};

        \draw (2.75,2.5) rectangle (3.75,3.5);
        \node at (3.25,3) {$h_1$};
        \draw (4,2.5) rectangle (5,3.5);
        \node at (4.5,3) {$h_2$};
        \draw (5.25,2.5) rectangle (6.25,3.5);
        \node at (5.75,3) {$h_3$};

        \draw (3,2) -- (3,2.5);
        \draw (6,2) -- (6,2.5);
        
        \draw (3,2.375) -- (4.25,2.375) -- (4.25,2.5);
        \node[fill=black, circle, inner sep=1pt] at (3,2.375) {};
        \draw (4.25,2.375) -- (5.5,2.375) -- (5.5,2.5);
        \node[fill=black, circle, inner sep=1pt] at (4.25,2.375) {};
        \draw (6,2.25) -- (4.75,2.25) -- (4.75,2.5);
        \node[fill=black, circle, inner sep=1pt] at (6,2.25) {};
        \draw (4.75,2.25) -- (3.5,2.25) -- (3.5,2.5);
        \node[fill=black, circle, inner sep=1pt] at (4.75,2.25) {};
        \draw (3.25,3.5) -- (3.25,4);
        \draw (4.5,3.5) -- (4.5,4);
        \draw (5.75,3.5) -- (5.75,4);

        \draw [decorate,decoration={brace,mirror,amplitude=5pt}] (0.25,1.5) -- (2,1.5);
        \node at (1.125,1) {$\vec{x}_\text{BS}$};
        \draw [decorate,decoration={brace,mirror,amplitude=5pt}] (3,1.5) -- (6,1.5);
        \node at (4.5,1) {$\vec{x}_\text{FS}$};
\end{tikzpicture}}
    \caption{Illustration of rephrased Ashenhurst–Curtis decomposition. A four-variable function is decomposed into FS functions $\vec{h}=\{h_1,h_2,h_3\}$ and BS functions $\vec{g}=\{g_1,g_2\}$. The top function $l$ is implemented as a multiplexer, using $\vec{g}$ as select signals to choose among $\vec{h}$.}
    \label{fig:ACD}
\end{figure}

Following recent advances in scalable ACD realization~\cite{calvino2025decomp}, we interpret $l$ as a \emph{multiplexer}~(MUX) whose data inputs are the FS functions $\vec{h}$ over $\vec{x}_{\text{FS}}$ and whose select lines are the BS functions $\vec{g}$ together with the SS variables.
Since BS and SS variables play symmetric roles in the selection logic, we merge them into a single set of \emph{selector variables}, still denoted $\vec{x}_{\text{BS}}$ for simplicity.

Under this consolidation, the ACD form becomes:
\begin{equation*}
    f(\vec{x}) = l\big(\vec{g}(\vec{x}_{\text{BS}}),\ \vec{h}(\vec{x}_{\text{FS}})\big),
\end{equation*}
where $l$ is explicitly realized as the top-level MUX.
This equivalent formulation preserves the generality of ACD while simplifying the variable-partitioning problem -- a key advantage for the JMC-oriented synthesis engine developed in this section.

An example is shown in Fig.~\ref{fig:ACD}, where a four-variable function is decomposed into three FS functions $\vec{h}=\{h_1,h_2,h_3\}$ and two BS functions $\vec{g}=\{g_1,g_2\}$.

Guided by this formulation, our synthesis engine is structured into three stages:
(1) partition variables and derive the FS functions~($\vec{h}$);
(2) assign encodings to the FS functions and construct the BS functions~($\vec{g}$);
(3) synthesize the top MUX~($l$).
In the remainder of this subsection, we describe the design for the first two stages.
The construction of the MUX is deferred to the next subsection.

\subsubsection{Variable Partitioning and Free-Set Function Construction}
\label{subsubsec:fs_funcs}

\begin{algorithm}[b]
    \DontPrintSemicolon
    \caption{Ownership-Aware Variable Partitioning}
    \label{alg:partition_fs}
    \KwIn{Cut $c$ with leaf set $L$,\\ownership map $\mathsf{own}:L\mapsto\mathcal{P}$}
    \KwOut{Set of candidate partitions $\mathcal{S}$}
    \ForEach{$\mathcal{P}_i \in \mathcal{P}$}
    {
        $L_i \gets \{\,\ell\in L \mid \mathsf{own}(\ell)=\mathcal{P}_i\,\}$
    }
    
    $\mathcal{S} \gets \emptyset$\;
    \ForEach{$\mathcal{P}_i \in \mathcal{P}$}
    {
        $\vec{x}_{\text{FS}} \gets L_i$\;
        $\vec{x}_{\text{BS}} \gets (L \setminus L_i)$\;
        \If{$|\vec{x}_{\text{FS}}| > 1$}
        {
            add $(\vec{x}_{\text{BS}},\vec{x}_{\text{FS}})$ to $\mathcal{S}$\;
        }   
    }
    \Return{$\mathcal{S}$}
\end{algorithm}

Our variable partitioning procedure is described in Algorithm~\ref{alg:partition_fs}.
Given a cut $c$ with leaves $L$ and its function, we first separate exclusively owned leaves by party (lines 1-2).
This explicitly identifies signals that can be assigned to a single party’s local computation, and therefore, on the FS side.

We then construct ownership-aware candidate partitions by placing all leaves of a single party into the FS and all remaining leaves into the BS (lines 3-8).
Partitions with $|\vec{x}_{\text{FS}}| \leq 1$ are discarded, since they offer negligible local-computation leverage (lines 7-8).

Once the FS set $\vec{x}_{\text{FS}}$ is determined, the corresponding FS functions are uniquely fixed.
These functions are obtained by enumerating all value assignments of the BS variables: for each assignment, the restricted function defines one FS function.
Collecting the distinct outcomes yields the FS function set $\vec{h}$.
This process is most naturally visualized through the truth table representation.

\begin{figure}[t]
    \centering
    \begin{tikzpicture}

\node (table1) at (0, 0) {
    \begin{tabular}{ccc|c}
        $b_2$ & $b_1$ & $a_1$ & $f$ \\
        \hline
        0 & 0 & 0 & 1 \\
        0 & 0 & 1 & 1 \\
        0 & 1 & 0 & 0 \\
        0 & 1 & 1 & 1 \\
        1 & 0 & 0 & 0 \\
        1 & 0 & 1 & 0 \\
        1 & 1 & 0 & 0 \\
        1 & 1 & 1 & 0 \\
        \hline
    \end{tabular}
};

\node (table2) at (5, 0) {
    \begin{tabular}{c|cc|cc}
        $\vec{x}_{\text{BS}}$ & \multicolumn{2}{c|}{$\vec{x}_{\text{FS}}$} & \\
        $a_1$ & $b_2$ & $b_1$ & $f$ \\
        \cline{1-4}
        0 & 0 & 0 & 1 & \multirow{4}{*}{$h_1$}\\
        0 & 0 & 1 & 0 & \\
        0 & 1 & 0 & 0 & \\
        0 & 1 & 1 & 0 & \\
        \cline{1-4}
        1 & 0 & 0 & 1 & \multirow{4}{*}{$h_2$}\\
        1 & 0 & 1 & 1 & \\
        1 & 1 & 0 & 0 & \\
        1 & 1 & 1 & 0 & \\
        \cline{1-4}
    \end{tabular}
};

\node[anchor=base] (arr) at (2.2,0) {$\xrightarrow[\text{variables}]{\text{reorder}}$};

\end{tikzpicture}
    \caption{Derivation of FS functions from the truth table under the partition $\vec{x}_{\text{FS}}=\{b_1, b_2\}$, $\vec{x}_{\text{BS}}=\{a_1\}.$ By enumerating the values of the BS variable $a_1$ and reordering the truth table accordingly, two distinct FS functions are identified.}
    \label{fig:derive_fs_funcs}
\end{figure}

Consider the example in Fig.~\ref{fig:limitation_of_mc}.
Among the three primary inputs, $a_1$ is exclusively owned by party $\mathcal{A}$, while $b_1$ and $b_2$ are exclusively owned by party $\mathcal{B}$.
According to Algorithm~\ref{alg:partition_fs}, this leads to a single valid partition: $\vec{x}_{\text{FS}}=\{b_1, b_2\}$ and $\vec{x}_{\text{BS}}=\{a_1\}$.
As shown in Fig.~\ref{fig:derive_fs_funcs}, reordering the variables in the truth table reveals two distinct FS functions, namely $\vec{h}=\{h_1, h_2\}$, with truth tables \texttt{\#1} and \texttt{\#3}.
These correspond to the Boolean expressions $h_1=\neg b_1\neg b_2$ and $h_2=\neg b_2.$

\subsubsection{Encoding and Bound-Set Function Derivation}
\label{subsubsec:bs_funcs}

The number of FS functions $|\vec{h}|$ directly determines the number of BS functions required to distinguish them.
In principle, $\log_2|\vec{h}|\leq|\vec{g}|\leq|\vec{h}|$.
However, in an MPC context, the computation of BS functions typically involves multiple parties, thereby contributing to the jointly derived computation.
To reduce this cost, we enforce the constraint
\begin{equation*}
    |\vec{g}| = \lceil\log_2|\vec{h}|\rceil,
\end{equation*}
ensuring that the number of BS functions is minimized while still preserving the ability to encode all FS functions uniquely.


This formulation naturally connects the problem to an \emph{encoding} task: each FS function must be assigned at least one $\lceil \log_2|\vec{h}|\rceil$-bit code that specifies its selection by the BS functions at the input of the top multiplexer $l$.
In the simple case previously introduced in Fig.~\ref{fig:derive_fs_funcs}, two FS functions are present, i.e., $|\vec{h}|=2$.
The encoding is straightforward: one function is assigned code $0$ and the other code $1$.
This implies a single BS function $g_1$, implemented directly over the BS variable $a_1$.
Depending on the assignment, $g_1$ may realize either the identity $g_1=a_1$ or the negation $g_1=\neg a_1$, both of which yield equivalent garbling costs.

While the encoding is trivial for small $|\vec{h}|$, the general case involves a significantly larger design space.
Each FS function must be mapped to a codeword such that all functions remain distinguishable, and the induced BS functions are minimized in both number and support size.
This can be interpreted as a \emph{constrained encoding} problem and solved via a two-step \emph{encoding-minimization} process~\cite{calvino2025decomp}:
\begin{enumerate}
    \item \textbf{Candidate code generation}: Each FS function is associated with at least one codeword of length $\lceil \log_2|\vec{h}|\rceil$.
    Codewords are selected so that different FS functions are mapped to distinct encodings, guaranteeing that they can be separated by the BS signals.
    \item \textbf{Support minimization}: For each bit position of the codewords, the induced Boolean function over the BS variables defines a candidate BS function.
    Since multiple encodings may be valid for a given FS function, the assignment is optimized so that the resulting BS functions have minimal support, i.e., they depend on as few BS variables as possible.
    This reduces the complexity of the BS side of the decomposition, thereby limiting the amount of joint computation required.
\end{enumerate}
The outcome of this procedure is a strict encoding, where each FS function is assigned a unique code, leading to a decomposition that balances distinguishability with minimal joint computation.  
The codes correspond to a compact set of BS functions $\vec{g}$.

\subsubsection{Synthesis of Derived BS and FS Functions}
\label{subsubsec:bs_fs_synthesis}

At this point, both the set of BS functions $\vec{g}$ and the set of FS functions $\vec{h}$ have been derived, and the next task is to implement them as XAGs.
For this purpose, we \rev{use} a near-MC-minimum XAG synthesizer based on the \rev{exact} SAT formulation introduced in~\cite{soeken2020determining}.
\rev{Since solving SAT instances on the fly would be too slow inside a peephole loop, the exact results are precomputed: every function of up to six variables is reduced to the representative of its spectral equivalence class (which preserves MC), and a database stores an MC-optimal XAG per class, so that each BS or FS output is obtained by canonicalization and lookup rather than by solving.
Both the BS and FS functions are, however, \emph{multi-output} functions over the BS or FS variables, respectively, and the database provides optimality per output only: outputs are synthesized one at a time and merged by structural hashing, so AND nodes are shared only when they coincide structurally.
To quantify the resulting gap, we extended the formulation of~\cite{soeken2020determining} to exact multi-output synthesis, where all outputs share one pool of AND gates, and solved every distinct BS and FS instance that occurs during the optimization of the benchmarks in \S\ref{sec:eval}.
For the $353$ BS instances -- whose AND gates are jointly evaluated -- the per-output lookup uses $476$ AND gates against an exact optimum of $472$ ($<1\%$ gap) and is exactly optimal in $98.9\%$ of the instances.
For the $1997$ FS instances, it is optimal in $90.9\%$ of the cases and uses $1320$ against $1095$ AND gates; this FS-side surplus is \emph{purely local} and it never enters the resulting garbled circuit.
Per-output lookup thus matches exact multi-output synthesis where it matters for JMC, at negligible runtime.}

Why measure BS/FS quality by MC? The rationale lies in variable ownership.
As discussed in \S~\ref{subsubsec:fs_funcs}, BS variables are typically \emph{jointly derived}, so most BS functions are evaluated jointly and their MC coincides with their JMC, making MC minimization directly relevant.
FS variables are, by construction, exclusively owned by a single party, so FS functions are purely local; nonetheless we synthesize them with the same incremental near-MC-minimum engine for consistency.

With the BS and FS functions synthesized, the only remaining task to fully reconstruct the original function $f$ is the synthesis of the top multiplexer $l$, which uses the outputs of the $|\vec{g}|$ BS functions to select among the outputs of the $|\vec{h}|$ FS functions.
This final step is detailed in the next subsection.

\subsection{Low-MC Multiplexer Construction}
\label{subsec:low_mc_mux}

\paragraph{Why target MC-optimality}
As established in \S~\ref{subsubsec:bs_fs_synthesis}, the data signals (outputs of FS functions) belong to a single party, whereas the select signals (outputs of BS functions) are typically \emph{jointly derived}.
In standard MUX realizations (e.g., $\neg\mathsf{sel}\cdot\mathsf{in_0}+\mathsf{sel}\cdot\mathsf{in_1}$), data and select entangle from the first logic level, which makes it reasonable to treat the entire MUX as joint computation.
Consequently, for MUX synthesis we have $\mathrm{JMC}=\mathrm{MC}$, and pursuing MC-optimal (or provably low-MC) constructions directly reduces the GC cost.

\paragraph{Base case: $2$-to-$1$ MUX with one AND (MC/JMC $=1$)}
Using the SAT-based MC-minimal XAG synthesis for single-output functions~\cite{soeken2020determining}, a $2$-to-$1$ MUX requires exactly one AND: $\mathrm{MUX}(\mathsf{sel}, \mathsf{in}_0, \mathsf{in}_1) = \mathsf{sel}\ ?\ \mathsf{in}_1 : \mathsf{in}_0 = [(\mathsf{in}_0 \oplus \mathsf{in}_1) \wedge \mathsf{sel}] \oplus \mathsf{in}_0$; intuitively, if the data inputs are equal the output ignores $\mathsf{sel}$, and if they differ it toggles by $\mathsf{sel}$ relative to $\mathsf{in}_0$.

\paragraph{General case: $n$-to-$1$ MUX via binary composition}
A well-known decomposition builds an $n$-to-$1$ MUX as a full binary tree of $2$-to-$1$ MUXes, requiring $(n{-}1)$ such nodes~\footnote{The binary-tree composition of multiplexers is standard in digital design; any fan-in-$n$ MUX can be realized by $(n{-}1)$ cascaded $2$-to-$1$ MUXes organized as a balanced tree.}.  
Substituting the one-AND realization for each $2$-to-$1$ MUX yields an XAG with $\mathrm{MC}=\mathrm{JMC}=n-1$ for the top MUX; the construction is deterministic and adds negligible synthesis time.

\paragraph{Implication for our ACD-based flow}
Since the top MUX $l$ selects among $|\vec{h}|$ FS outputs using $|\vec{g}|=\lceil \log_2 |\vec{h}|\rceil$ select signals, the above construction gives $\mathrm{JMC}(l)=|\vec{h}|-1.$
Applying this to the running example yields $\mathrm{JMC}(l)=1$ and reproduces Fig.~\ref{fig:example_opt}, achieving a $50\%$ reduction in JMC compared to the implementation in Fig.~\ref{fig:example_orig}.

In summary, these three techniques -- (i) joint compute-oriented decomposition, (ii) near-MC-minimal synthesis of the derived BS and FS functions, and (iii) low-MC MUX construction -- constitute the sub-network resynthesis engine invoked in line~7 of Algorithm~\ref{alg:frame}, the computational core that replaces sub-networks with optimized XAGs minimizing jointly derived computation.

\section{Experimental Evaluation}
\label{sec:eval}
We conducted a series of experiments on representative benchmarks to evaluate the effectiveness of the proposed ownership-aware optimization framework.
The goal is to assess how our approach uncovers circuit-level opportunities for reducing jointly evaluated computation, complementing and extending beyond conventional MC-driven reduction.

\subsection{Experimental Setup}
All experiments were conducted on an Apple M1 Max machine with 32\,GB RAM.
All benchmarks were first preprocessed using high-effort MC-reduction techniques~\cite{testa2020mc,liu2022dontcare}.
The framework (Algo.~\ref{alg:frame}) is built atop the \texttt{mockturtle} logic synthesis library~\cite{soeken2022mockturtle}, and is available with all reported data at \url{github.com/MingfeiYu/jmc-synthesis}.
For each benchmark, the optimization iteratively applies the framework until no further improvement emerges.
The cut-size bound $\kappa$ is empirically set to $8$, offering a favorable balance between optimization granularity and runtime\rev{; the sensitivity to this choice is investigated and reported}.
\rev{Cuts are enumerated with \texttt{mockturtle}'s priority-cut enumeration, which keeps at most $8$ cuts per node.}

The evaluation takes two complementary perspectives: a circuit-level analysis that quantifies how ownership-aware optimization reshapes Boolean circuits, isolating the reduction of jointly evaluated computation from system-level factors, and a system-level evaluation of the optimized circuits in a full MPC framework, \texttt{MP-SPDZ}~\cite{mp-spdz}, which measures how the restructuring translates into communication and latency.


\subsection{Benchmarks}

We evaluate our method on 12 circuits drawn from three benchmark suites and organize them into two categories.

(1) \emph{Arithmetic and comparative functions}:
This category includes seven benchmarks performing numerical or relational operations: 128-bit integer addition (\texttt{INT-add}), 64-bit division (\texttt{INT-div}), 64-bit multiplication (\texttt{INT-mul}), 64-bit floating-point addition (\texttt{FP-add}), 64-bit floating-point division (\texttt{FP-div}), 64-bit floating-point multiplication (\texttt{FP-mul}), and 64-bit floating-point equality (\texttt{FP-eq}).
The integer circuits come from the EPFL suite~\cite{amaru2015epfl}, and the floating-point circuits from Nigel Smart’s MPC Circuits collection~\cite{mpc_circuits}.
All compute a relation over two operands; accordingly, each operand is assigned to a different party, yielding a natural two-party ownership model.

(2) \emph{Cryptographic and general-purpose kernels}:
This category comprises five circuits from the \texttt{TinyGarble} suite~\cite{songhori2015tinygarble}: \texttt{AES-128}, \texttt{SHA3}, 32-bit \texttt{Hamming} distance, 32-bit \texttt{Encoder}, and the \texttt{LiteMIPS} processor.
For these benchmarks, we adopt the original input-ownership assignments provided with the suite.

Although all benchmarks instantiate two-party computations, neither JMC nor the proposed framework is limited to two parties (\S~\ref{sec:conclusion}).

\subsection{Circuit-Level Evaluation}
\label{subsec:circuit_level_eval}

We report, for the baseline XAG after MC reduction and after our framework, the total number of AND gates (`\#ANDs') and the subset identified as jointly derived under ownership analysis (`\#Joint ANDs', \S\ref{subsec:ownership}), together with the relative reduction in garbled AND gates, `\text{Red.}'$= \big[\#\text{ANDs (w/o opt.)} - \#\text{Joint ANDs (w/ opt.)}\big] / \#\text{ANDs (w/o opt.)}$.

\begin{table}[t]
    \centering
    \resizebox{0.7\linewidth}{!}{\begin{tabular}{lrrrrr}
    \toprule[1.2pt]
    Benchmark &
    \makecell{\#ANDs\\(w/o opt.)} &
    \makecell{\#Joint ANDs\\(w/o opt.)} &
    \makecell{\#ANDs\\(w/ opt.)} &
    \makecell{\#Joint ANDs\\(w/ opt.)} &
    \makecell{Red.\\{} [\%]} \\
    \midrule
    \texttt{INT-add} & 128     & 128     & 128     & 128     & 0.00  \\
    \texttt{INT-div} & 5\,291  & 5\,205  & 5\,265  & 5\,133  & 2.99  \\
    \texttt{INT-mul} & 7\,653  & 7\,410  & 7\,656  & 7\,376  & 3.62  \\
    \texttt{FP-add}  & 5\,141  & 4\,938  & 5\,038  & 4\,824  & 6.17  \\
    \texttt{FP-div}  & 48\,408 & 44\,834 & 48\,149 & 44\,564 & 7.94  \\
    \texttt{FP-mul}  & 16\,684 & 13\,113 & 16\,439 & 12\,859 & 22.93 \\
    \texttt{FP-eq}   & 315     & 191     & 315     & 97      & 69.21 \\
    \bottomrule[1.2pt]
\end{tabular}
}
    \caption{Results on arithmetic and comparative benchmarks.} 
\label{tab:results_arith}
\end{table}

\paragraph{Analysis of results on arithmetic and comparative benchmarks}
The first group of results (Table~\ref{tab:results_arith}) exhibits two clear patterns:
(1) Floating-point operators provide considerably more opportunities for reducing joint computation than integer ones; and
(2) Floating-point comparison benefits most from our optimization framework.

While integer operators are structurally dominated by joint dependencies, floating-point operators differ fundamentally.
Despite the large circuit size, only a core portion of the computation, such as mantissa addition for \texttt{FP-add} or partial-product accumulation for \texttt{FP-mul}, genuinely mixes the two operands; much of the surrounding logic (e.g., extracting fields, preparing operands, post-processing results) acts on each operand independently.
Our framework leverages this structure to isolate local regions and shrink the jointly derived portion.
This yields meaningful reductions for \texttt{FP-add}, \texttt{FP-div}, and \texttt{FP-mul}, while integer operators remain comparatively rigid.

\texttt{FP-eq} exhibits the largest reduction in this group, because its behavior is driven primarily by \emph{case analysis} rather than heavy two-operand arithmetic.
Many outcomes, such as handling zeros, infinities, or NaNs, depend on only one operand, and even the remaining comparison logic is implemented with lightweight predicates.
As a result, most of the computation is local, and the small joint core is aggressively compressed by our optimizer.

\begin{table}[b]
    \centering
    \resizebox{0.7\linewidth}{!}{\begin{tabular}{lrrrrr}
    \toprule[1.2pt]
    Benchmark &
    \makecell{\#ANDs\\(w/o opt.)} &
    \makecell{\#Joint ANDs\\(w/o opt.)} &
    \makecell{\#ANDs\\(w/ opt.)} &
    \makecell{\#Joint ANDs\\(w/ opt.)} &
    \makecell{Red.\\{}[\%]} \\
    \midrule
    \texttt{AES-128}    & 6\,446  & 5\,153 & 6\,446 & 5\,153 & 20.06 \\
    \texttt{SHA3}       & 1\,600  & 1\,600 & 1\,600 & 1\,600 & 0.00  \\
    \texttt{Hamming}    & 48      & 48     & 48     & 48     & 0.00  \\
    \texttt{Encoder}    & 53      & 10     & 53     & 6      & 88.68 \\
    \texttt{LiteMIPS}   & 11\,465 & 9\,380 & 9\,133 & 6\,965 & 39.25 \\
    \bottomrule[1.2pt]
\end{tabular}}
    \caption{Results on crypto and general-purpose kernels.} 
    \label{tab:results_tinygarble}
\end{table}

\paragraph{Analysis of results on cryptographic and general-purpose kernels}
The cryptographic benchmarks exhibit limited potential for further joint-computation minimization, although \texttt{AES-128} and \texttt{SHA3} behave differently under ownership analysis.
For \texttt{AES-128}, a nontrivial portion of the circuit is already identified as locally derivable, corresponding primarily to its linear layers, while the nonlinear computation is concentrated in the S-boxes~\cite{nist:aes}.
In contrast, \texttt{SHA3} exposes essentially no locally derivable AND gates, as its nonlinear $\chi$ step immediately mixes signals that are already jointly derived~\cite{nist:sha3}.
Despite this difference, our optimizer finds no further reduction opportunity in either benchmark: their nonlinear regions are isolated and already highly optimized, and the signals entering them are predominantly jointly derived, leaving no structural flexibility to push the local/joint boundary deeper.
Ownership analysis can thus still reveal existing locality, as in \texttt{AES-128}, but restructuring is ineffective because the remaining nonlinear kernels are intrinsically joint.

The general-purpose kernels (\texttt{Hamming}, \texttt{Encoder}, \texttt{LiteMIPS}) contain richer mixtures of datapath and control logic, and their locality properties vary significantly.
\texttt{Hamming} distance entangles the two inputs immediately through bitwise XOR, eliminating any chance for local computation.
In contrast, the \texttt{LiteMIPS} processor exposes substantial regions of operand-local computation, such as instruction decoding and control steering.
In the MPC setting, this benchmark models a scenario where one party provides the program (instruction stream), while the other supplies the data and architectural state, enabling portions of the execution to remain locally computable.
This structure translate into a $39.25\%$ reduction in garbling cost: $18.19\%$ comes from incorporating ownership awareness, and an additional $21.06\%$ is achieved through the optimization framework.

\paragraph{Discussion: JMC vs. MC}
Although our objective is to reduce jointly derived computation, JMC reductions are sometimes accompanied by reductions in conventional MC -- yet our approach is not a variant of MC minimization.
When both improve, the JMC gain is typically larger (e.g., \texttt{INT-div} reduces MC by $26$ but JMC by $72$), showing the dominant gains come from eliminating joint computation rather than overall gate count.
Conversely, JMC can fall while MC stays fixed or even rises: \texttt{FP-eq} reduces JMC with unchanged MC, and \texttt{INT-mul} reduces JMC despite a slight MC increase, so minimizing JMC may require structures deemed suboptimal under MC.
Thus MC- and JMC-driven optimization explore overlapping but distinct regions of the design space: where MC is not yet minimal our method may reduce both, but where MC is already aggressively minimized it exploits an orthogonal dimension -- maximizing local computation -- confirming that it is complementary to prior work and exposes opportunities beyond traditional MC minimization.

\paragraph{\rev{Sensitivity to the cut-size bound $\kappa$}}
\label{par:kappa_sensitivity}

\rev{Table~\ref{tab:results_kappa_sweep} reports the joint ANDs after optimization for $\kappa\in\{4,6,8,10\}$ on the benchmarks whose result depends on $\kappa$ (the others are unaffected at every $\kappa$).
First, $\kappa$ is bounded from above by the backend: the database of MC-optimal XAGs (\S\ref{subsubsec:bs_fs_synthesis}) covers at most six variables, so BS functions of larger support cannot be implemented.
At $\kappa=8$ this never rejects a candidate; at $\kappa=10$ it rejects $76\,800$ candidates on \texttt{LiteMIPS} (whose result degrades to $7\,157$) and $216\,183$ on \texttt{INT-mul}, and \texttt{INT-div}/\texttt{FP-div} exceed the truth-table width of the decomposition engine; $\kappa=8$ is thus the largest value the backend exploits fully.
Second, results are not monotone in $\kappa$: $\kappa=4$ is best on \texttt{FP-add} ($4\,821$), \texttt{Encoder} ($4$), and \texttt{LiteMIPS} ($6\,651$), $\kappa=6$ on \texttt{INT-div} ($5\,125$), while all other differences, except on \texttt{FP-eq}, stay below $1\%$.
The cause is the priority-cut budget rather than the scope: at $\kappa=8$ on \texttt{LiteMIPS}, we observed that $98\%$ of the accepted rewrites use cuts of $3$--$4$ leaves and none uses $7$--$8$, whereas large cuts occupy most of the $8$ slots per node and evict profitable small cuts.
Raising the budget to $15$ cuts at $\kappa=8$ (column $8^{\ast}$) yields the best result on every benchmark, except on \texttt{Encoder}, at $2.9\times$ the total synthesis time;
we keep $\kappa=8$ with the default budget as reference configuration, noting that a larger scope pays off only when profitable small cuts are not crowded out.}

\begin{table}[t]
    \centering
    \resizebox{0.65\linewidth}{!}{{\small\rev{\begin{tabular}{lrrrrr}
    \toprule[1.2pt]
    Benchmark & $\kappa=4$ & $\kappa=6$ & $\kappa=8$ & $\kappa=8^{\ast}$ & $\kappa=10$ \\
    \midrule
    \texttt{INT-div}  & 5\,164  & 5\,125  & 5\,133  & \textbf{5\,115}  & --      \\
    \texttt{INT-mul}  & 7\,399  & 7\,377  & 7\,376  & \textbf{7\,343}  & 7\,375  \\
    \texttt{FP-add}   & \textbf{4\,821}  & 4\,826  & 4\,824  & \textbf{4\,821}  & 4\,825  \\
    \texttt{FP-div}   & 44\,592 & 44\,564 & 44\,564 & \textbf{44\,528} & --      \\
    \texttt{FP-mul}   & 12\,962 & 12\,859 & 12\,859 & \textbf{12\,858} & 12\,859 \\
    \texttt{FP-eq}    & 127     & 97      & 97      & \textbf{95}      & 97      \\
    \texttt{Encoder}  & \textbf{4}       & 6       & 6       & 6       & 6       \\
    \texttt{LiteMIPS} & 6\,651  & 6\,936  & 6\,965  & \textbf{6\,638}  & 7\,157  \\
    \midrule
    Total synth.\ time [s] & 38 & 132 & 271 & 793 & -- \\
    \bottomrule[1.2pt]
\end{tabular}}
}}
    \caption{\rev{Joint ANDs achieved after optimization with various cut-size bound $\kappa$ ($8^{\ast}$: $\kappa=8$ with $15$ instead of the default $8$ priority cuts per node; --: unsupported by the backend; best per row is highlighted in bold).}}
    \label{tab:results_kappa_sweep}
\end{table}

\paragraph{\rev{Cost of optimization}}
\label{par:cost_of_opt}
\rev{The right part of Table~\ref{tab:results_tradeoff} reports the offline synthesis time and the structural side effects for the benchmarks of \S\ref{subsec:system_level_eval}.
Synthesis is a one-time cost of at most $100$\,s per benchmark.
JMC minimization admits additional local logic when it displaces joint computation, but this overhead is small: the total AND count does not increase for $8$ of the $9$ circuits and grows by three gates for \texttt{INT-mul}.
Depth, which determines the round count of GMW-style protocols~\cite{schneider2013lowdepth,demmler2015aby} and the critical path of parallel GC evaluation~\cite{buescher2015parallel}, is essentially preserved: the AND depth changes only on three circuits (\texttt{INT-div} $2\,244\to2\,232$, \texttt{FP-eq} $9\to10$, \texttt{Encoder} $4\to5$), and total depth grows noticeably only on the two smallest circuits, where a single ACD-based rewrite, whose select tree adds $\lceil \log_2 |\vec{h}|\rceil$ levels, spans a large fraction of the logic.}

\subsection{System-Level Evaluation in MPC}
\label{subsec:system_level_eval}

In this experiment, we focus on the subset of benchmarks from Tables~\ref{tab:results_arith} and \ref{tab:results_tinygarble} for which our approach yields a circuit with reduced JMC.
Benchmarks where the optimized circuit is identical to the baseline are omitted, as they do not provide additional insight at the system level.
All experiments are repeated $100$ times, and the reported results correspond to the average across runs, in order to mitigate the impact of system-level noise and obtain stable measurements.

We report the following metrics, summarized in Table~\ref{tab:results_mp-spdz}: `\#Inputs (G/E)' denotes the number of input bits held by the garbler (G) and the evaluator (E), respectively.
As the two roles can be assigned freely in our semi-honest setting, we make the party with fewer input bits the evaluator, to minimize input-label delivery.
\rev{Every frontier signal exposed by the optimization (\S\ref{subsubsec:node_ownership}) is fed to the garbled circuit as an ordinary private input of its owner: signals computed by the evaluator enter through OT exactly like its original input bits, and signals computed by the garbler are encoded directly as wire labels; no locally derived value is ever fixed inside the garbled tables.
Restructuring thus yields a circuit $C'$ with $C(x_A,x_B)=C'(L_A(x_A),L_B(x_B))$, where $L_i$ depends only on party $i$'s input; standard GC security applies to $C'$, and each party's local computation is simulatable from its own input.}
`Oblivious Transfer Comm. \& Time' report the data sent from the garbler to the evaluator and the wall-clock time of the OT phase; `OT Label' isolates the input-label delivery stage.
`Garbled Gates Comm. \& Eval.' denote the total size of the transmitted garbled tables and the GC evaluation time.
`Total Comm.' is the sum of `OT Comm.' and `GG Comm.'.

\begin{table*}[t]
    \centering
    \resizebox{\linewidth}{!}{\begin{tabular}{lcccccccc}
    \toprule[1.2pt]
    \multirow{2}{*}{Benchmark} & \multirow{2}{*}{\#Inputs (G/E)} & \multicolumn{3}{c}{Oblivious Transfer~(OT)} & \multicolumn{2}{c}{Garbled Gates~(GG)} & \multirow{2}{*}{Total Comm. [KB]} \\
    \cmidrule(lr){3-5} \cmidrule(lr){6-7}
     & & Label [ms] & Comm. [KB] & Time [ms] & Comm. [KB] & Eval. [ms] & \\
    \midrule
    \texttt{INT-div} (base) & 64/64 & 0.028 & 5.248 & 17.062 & 169.440 & 3.360 & 174.688 \\
    \texttt{INT-div} (opt.) & 189/91 & 0.030 ($\uparrow$ 7.14\%) & 5.680 ($\uparrow$ 8.23\%) & 16.899 ($\downarrow$ 0.96\%) & 164.384 ($\downarrow$ 2.98\%) & 3.295 ($\downarrow$ 1.93\%) & 170.064 ($\downarrow$ 2.65\%) \\
    \midrule
    \texttt{INT-mul} (base) & 64/64 & 0.028 & 5.248 & 16.648 & 245.024 & 0.493 & 250.272 \\
    \texttt{INT-mul} (opt.) & 254/238 & 0.032 ($\uparrow$ 14.29\%) & 8.032 ($\uparrow$ 53.05\%) & 16.248 ($\downarrow$ 2.40\%) & 236.160 ($\downarrow$ 3.62\%) & 0.438 ($\downarrow$ 11.16\%) & 244.192 ($\downarrow$ 2.43\%) \\
    \midrule
    \texttt{FP-add} (base) & 64/64 & 0.027 & 5.248 & 18.107 & 164.576 & 0.431 & 169.824 \\
    \texttt{FP-add} (opt.) & 125/124 & 0.026 ($\downarrow$ 3.70\%) & 6.208 ($\uparrow$ 18.29\%) & 17.997 ($\downarrow$ 0.61\%) & 154.432 ($\downarrow$ 6.16\%) & 0.441 ($\uparrow$ 2.32\%) & 160.640 ($\downarrow$ 5.41\%) \\
    \midrule
    \texttt{FP-div} (base) & 64/64 & 0.026 & 5.248 & 16.850 & 1549.120 & 5.994 & 1554.368 \\
    \texttt{FP-div} (opt.) & 183/149 & 0.029 ($\uparrow$ 11.54\%) & 6.608 ($\uparrow$ 25.91\%) & 16.326 ($\downarrow$ 3.11\%) & 1426.110 ($\downarrow$ 7.94\%) & 5.921 ($\downarrow$ 1.22\%) & 1432.718 ($\downarrow$ 7.83\%) \\
    \midrule
    \texttt{FP-mul} (base) & 64/64 & 0.029 & 5.248 & 16.035 & 533.952 & 0.758 & 539.200 \\
    \texttt{FP-mul} (opt.) & 143/132 & 0.030 ($\uparrow$ 3.45\%) & 6.336 ($\uparrow$ 20.73\%) & 15.288 ($\downarrow$ 4.66\%) & 411.552 ($\downarrow$ 22.92\%) & 0.613 ($\downarrow$ 19.13\%) & 417.888 ($\downarrow$ 22.50\%) \\
    \midrule
    \texttt{FP-eq} (base) & 64/64 & 0.040 & 5.248 & 23.979 & 10.081 & 0.039 & 15.329 \\
    \texttt{FP-eq} (opt.) & 81/81 & 0.040 ($\uparrow$ 0.00\%) & 5.520 ($\uparrow$ 5.18\%) & 23.525 ($\downarrow$ 1.89\%) & 3.105 ($\downarrow$ 69.20\%) & 0.027 ($\downarrow$ 30.77\%) & 8.625 ($\downarrow$ 43.73\%) \\
    \midrule
    \texttt{AES-128} (base) & 128/128 & 0.031 & 6.272 & 16.337 & 206.400 & 0.187 & 212.672 \\
    \texttt{AES-128} (opt.) & 1408/128 & 0.030 ($\downarrow$ 3.23\%) & 6.272 ($\uparrow$ 0.00\%) & 16.211 ($\downarrow$ 0.77\%) & 165.024 ($\downarrow$ 20.05\%) & 0.166 ($\downarrow$ 11.23\%) & 171.296 ($\downarrow$ 19.46\%) \\
    \midrule
    \texttt{Encoder} (base) & 16/16 & 0.038 & 4.480 & 25.398 & 1.701 & 0.020 & 6.181 \\
    \texttt{Encoder} (opt.) & 5/5 & 0.035 ($\downarrow$ 7.89\%) & 4.304 ($\downarrow$ 3.93\%) & 24.275 ($\downarrow$ 4.42\%) & 0.196 ($\downarrow$ 88.48\%) & 0.016 ($\downarrow$ 20.00\%) & 4.500 ($\downarrow$ 27.20\%) \\
    \midrule
    \texttt{LiteMIPS} (base) & 2048/2048 & 0.068 & 36.992 & 15.719 & 374.046 & 0.373 & 411.038 \\
    \texttt{LiteMIPS} (opt.) & 3072/372 & 0.029 ($\downarrow$ 57.35\%) & 10.176 ($\downarrow$ 72.49\%) & 16.399 ($\uparrow$ 4.33\%) & 225.920 ($\downarrow$ 39.60\%) & 0.247 ($\downarrow$ 33.78\%) & 236.096 ($\downarrow$ 42.56\%) \\
    \bottomrule[1.2pt]
\end{tabular}
}
    \caption{End-to-end MPC evaluation, showing the impact of circuit restructuring on oblivious transfer and garbled-gate costs.} 
\label{tab:results_mp-spdz}
\end{table*}

\paragraph{Communication cost analysis}

In end-to-end MPC execution, communication arises from the OT phase and from the transmission of garbled gates (`OT Comm.' and `GG Comm.').

A notable effect of our optimization is that local precomputation may increase the number of input bits to the joint computation: the optimized circuit exposes additional derived signals that, together with the original inputs, form the interface to the joint computation.
This raises a concern that larger evaluator input size inflates input-label delivery and hence OT runtime.
The `OT Label' column isolates this cost, and the effect is barely observable.
Although label delivery scales linearly with evaluator input bits in theory, in practice it is dominated by a fixed per-call overhead of $10$--$25\,\mu$s: for \texttt{INT-mul} the evaluator inputs grow more than $3\times$ ($64$ vs.\ $238$) yet label time rises only $14.29\%$, and for \texttt{FP-add} ($64$ vs.\ $124$) it even decreases by $3.70\%$, as the additional labels amortize within the same system call.

The `OT Comm.' column tells the same story.
\texttt{MP-SPDZ}'s IKNP-style OT extension~\cite{otextension} sends a fixed Base OT setup ($\approx\!4.2$\,KB) plus $16$\,B per evaluator input bit; only the latter scales with input size, so even a substantial input increase adds only modest OT communication.

In contrast, the `GG Comm.' column exhibits a direct proportionality to the number of AND gates.
\texttt{MP-SPDZ} adopts the half-gate construction~\cite{zahur2015halfgate}, in which each AND gate is encoded using exactly two $128$-bit ciphertexts ($32$\,B), while XOR gates incur no cost, so removing one jointly evaluated AND gate deterministically saves $32$\,B of communication.

This asymmetry between OT and garbled-gate communication justifies our design philosophy: reducing JMC may add evaluator input bits, but each costs only $16$\,B of OT on top of a large constant base, whereas each eliminated AND gate saves a fixed $32$\,B.
Therefore, reducing jointly evaluated AND gates remains beneficial even when evaluator inputs increase, as observed across all benchmarks.
\rev{Table~\ref{tab:results_tradeoff} makes this trade-off explicit for every benchmark: the additional evaluator inputs, the resulting OT communication ($16$\,B each), the garbled ANDs saved, the garbled-gate communication ($32$\,B each), and the net change.
The garbled-gate saving exceeds the OT overhead by at least $3.2\times$ (\texttt{INT-mul)}) and typically by one to two orders of magnitude; for \texttt{Encoder} and \texttt{LiteMIPS} the optimized frontier is even narrower than the original input, so both terms decrease.}

\begin{table*}[b]
    \centering
    \resizebox{0.95\linewidth}{!}{\rev{\begin{tabular}{lrrrrrrcr}
    \toprule[1.2pt]
    \multirow{2}{*}{Benchmark} & \multirow{2}{*}{\makecell{$\Delta$Eval.\\inputs}} & \multirow{2}{*}{\makecell{$\Delta$OT\\comm.\ [B]}} & \multirow{2}{*}{$\Delta$\#GGs} & \multirow{2}{*}{\makecell{$\Delta$GG\\comm.\ [B]}} & \multirow{2}{*}{\makecell{$\Delta$Net\\comm.{}[B]}} & \multirow{2}{*}{$\Delta$\#ANDs} & \multicolumn{1}{c}{Depth / AND depth} & \multirow{2}{*}{\makecell{Synth.\\time [s]}} \\
    \cmidrule(lr){8-8}
    & & & & & & & base $\rightarrow$ opt. & \\
    \midrule
    \texttt{INT-div} & $\uparrow$ 27 & $\uparrow$ 432 & $\downarrow$ 158 & $\downarrow$ 5\,056 & $\downarrow$ 4\,624 & $\downarrow$ 26 & 5\,938/2\,244 $\rightarrow$ 5\,989/2\,232 & 100.3 \\
    \texttt{INT-mul} & $\uparrow$ 174 & $\uparrow$ 2\,784 & $\downarrow$ 277 & $\downarrow$ 8\,864 & $\downarrow$ 6\,080 & $\uparrow$ 3 & 328/149 (unchanged) & 14.0 \\
    \texttt{FP-add} & $\uparrow$ 60 & $\uparrow$ 960 & $\downarrow$ 317 & $\downarrow$ 10\,144 & $\downarrow$ 9\,184 & $\downarrow$ 103 & 553/234 (unchanged) & 17.8 \\
    \texttt{FP-div} & $\uparrow$ 85 & $\uparrow$ 1\,360 & $\downarrow$ 3\,844 & $\downarrow$ 123\,010 & $\downarrow$ 121\,650 & $\downarrow$ 259 & 8\,139/3\,550 (unchanged) & 78.7 \\
    \texttt{FP-mul} & $\uparrow$ 68 & $\uparrow$ 1\,088 & $\downarrow$ 3\,825 & $\downarrow$ 122\,400 & $\downarrow$ 121\,312 & $\downarrow$ 245 & 337/195 $\rightarrow$ 340/195 & 15.6 \\
    \texttt{FP-eq} & $\uparrow$ 17 & $\uparrow$ 272 & $\downarrow$ 218 & $\downarrow$ 6\,976 & $\downarrow$ 6\,704 & 0 & 9/9 $\rightarrow$ 14/10 & 5.7 \\
    \texttt{AES-128} & 0 & 0 & $\downarrow$ 1\,293 & $\downarrow$ 41\,376 & $\downarrow$ 41\,376 & 0 & 320/60 (unchanged) & 2.0 \\
    \texttt{Encoder} & $\downarrow$ 11 & $\downarrow$ 176 & $\downarrow$ 47 & $\downarrow$ 1\,505 & $\downarrow$ 1\,681 & 0 & 4/4 $\rightarrow$ 9/5 & 2.5 \\
    \texttt{LiteMIPS} & $\downarrow$ 1\,676 & $\downarrow$ 26\,816 & $\downarrow$ 4\,500 & $\downarrow$ 148\,126 & $\downarrow$ 174\,942 & $\downarrow$ 2\,332 & 67/50 (unchanged) & 29.2 \\
    \bottomrule[1.2pt]
\end{tabular}}

}
    \caption{\rev{Per-benchmark trade-off between extra evaluator inputs (OT) and saved garbled ANDs (GG), with synthesis time and depth.}}
    \label{tab:results_tradeoff}
\end{table*}

Moreover, higher OT communication does not translate into higher OT runtime: it stays within $15$--$27$\,ms, dominated by the fixed Base OT setup, with the $1$--$3$\,ms variation attributable to system noise.
(The evaluator-to-garbler IKNP correlation matrix~\cite{otextension} is a constant $\approx\!2.0$\,KB, which is unaffected by restructuring and omitted.)

\paragraph{Computation cost analysis}

The `GG Eval.' column captures the computational cost.
Theoretically, GC evaluation time scales with the number of AND gates (\S\ref{subsubsec:garbling_cost}), and `GG Eval.' generally follows the AND-gate reduction, but not strictly: \texttt{FP-add} shows a $2.32\%$ increase in evaluation time despite a $6.16\%$ reduction in AND gates.
We attribute such deviations to system-level effects of wall-clock measurement: evaluation time also reflects memory access and data movement, and when restructuring exposes more evaluator input wires (\texttt{FP-add}: $64\to124$), their externally delivered labels exhibit poorer cache locality, partially offsetting the garbled-gate savings; circuits with sub-millisecond runtime amplify such noise.
We therefore view `GG Eval.' as a useful but imperfect indicator, whereas the reduction in jointly evaluated AND gates (Tables~\ref{tab:results_arith} and \ref{tab:results_tinygarble}, and the `$\Delta$\#GGs' column in Table~\ref{tab:results_tradeoff}) is a deterministic, implementation-independent measure of the intrinsic computational savings.
\section{Positioning of Contributions}
\label{sec:position}
\rev{Relative to the prior art of \S\ref{sec:related}, our contribution is orthogonal along two axes.
Cryptographic optimizations (free-XOR, half-gates, three-halves) lower the cost \emph{per} garbled AND and MC-driven synthesis lowers the \emph{number} of ANDs; we lower the number of ANDs that must be garbled \emph{at all}, and compose with both (our baselines are MC-minimized and garbled with half-gates).
\texttt{SkipGate}/\texttt{ARM2GC}~\cite{songhori2019arm2gc} skip gates whose values are \emph{publicly} determined, and \texttt{Senate}~\cite{poddar2021senate} or \texttt{Marill}~\cite{rathee2024marill} exploit locality at the level of whole operations of one application; we exploit values known \emph{privately} to one party, automatically and at gate granularity, for arbitrary Boolean circuits.
Both ingredients matter: ownership analysis alone yields $18.19$ ($39.37$) percentage points of the reduction on \texttt{LiteMIPS} (\texttt{FP-eq}) and the resynthesis engine a further $21.06$ ($29.84$).}
\section{Conclusion}
\label{sec:conclusion}

This work optimizes garbled circuits by shifting the cost model from traditional \emph{multiplicative complexity}~(MC) to one that explicitly captures \emph{joint computation}.
By formalizing node ownership,
we established a refined metric -- \emph{joint multiplicative complexity}~(JMC) -- that more faithfully reflects actual GC-based MPC cost, and built a peephole optimization framework whose Ashenhurst–Curtis–decomposition-based synthesis engine resynthesizes sub-networks -- via ownership-aware variable partitioning and low-MC component implementations -- to reduce JMC while remaining scalable.
Our evaluation shows the method uncovers optimization opportunities entirely missed by conventional MC-based approaches, highlighting a new dimension in secure-computation design: maximizing local work while confining joint computation to where it is truly unavoidable.

\paragraph{\rev{Future directions}}
\rev{The ownership analysis of \S\ref{subsubsec:node_ownership} is already stated for $t$ parties; only the benchmarks and the cost model instantiated in \S\ref{sec:eval} are two-party.
What changes for $t>2$ is the cost model: generalizing $\mathsf{LD}_{\mathcal{P}_i}$ to sets of parties labels every signal with the minimal set $S\subseteq\mathcal{P}$ whose inputs it depends on, and the cost of a jointly derived AND then depends on $|S|$ and on the protocol -- in multi-party garbling, an AND costs $\Theta(t)$ ciphertexts irrespective of $S$, so JMC carries over directly; in secret-sharing-based protocols, a gate with $|S|<t$ may be evaluated by the sub-committee $S$, so its cost grows with $|S|$.
The decomposition engine only needs the partition into single-owner (free-set) and remaining (bound-set) variables and is unaffected; the extension thus amounts to a protocol-specific cost function over ownership sets.
We leave the exploration, together with malicious-security variants, where locally derived frontier inputs require consistency checks, to future work.}
\section*{Acknowledgments}
We thank the anonymous reviewers and our shepherd for their constructive feedback, which has substantially improved this paper.
Mingfei Yu gratefully acknowledges the support of CWI's internship programme for PhD students, and of the Swiss National Science Foundation through the PostDoc Mobility Fellowship P500-2\_235311.
Chenglu Jin is partially supported by the project CiCS of the research programme Gravitation, which is partly financed by the Dutch Research Council~(NWO) under grant 024.006.037.

\appendix

\bibliographystyle{alpha}
\bibliography{References}

\end{document}